\documentclass[preprint,aps,superscriptaddress,nofootinbib,longbibliography]{revtex4-1}
\usepackage[T1]{fontenc}
\usepackage[latin9]{inputenc}
\usepackage{amsmath}
\usepackage{graphicx}
\usepackage{babel}
\usepackage{url}
\usepackage{hyperref}
\usepackage{slashed}
\usepackage{amsthm}
\usepackage{xcolor}
\usepackage[normalem]{ulem}
\newtheorem*{theorem*}{Theorem}

\begin{document}
	\title{Theoretical Aspect of Nonunitarity in Neutrino Oscillation}
	
	\newcommand{\affUFABC}{Centro de Ci\^encias Naturais e Humanas\;\;\\
		Universidade Federal do ABC, 09.210-170,
		Santo Andr\'e, SP, Brazil}
	
	\author{Chee Sheng Fong}
	\email{sheng.fong@ufabc.edu.br}
	\affiliation{\affUFABC}

\begin{abstract}
Nonunitarity can arise in neutrino oscillation when the matrix with elements $\mathbf{U}_{\alpha i}$ which relate the neutrino flavor $\alpha$ and mass $i$ eigenstates is not unitary when sum over the kinematically accessible mass eigenstates or over the three Standard Model flavors. We review how high scale nonunitarity arises after integrating out new physics which is not
accessible in neutrino oscillation experiments. In particular, we stress that high scale unitarity violation is only \emph{apparent} and what happens is that the neutrino flavor states become nonorthogonal due to new physics. Since the flavor space is complete, unitarity has to be preserved in time evolution and that the probabilities of a flavor state oscillates to all possible flavor states always sum up to unity. We highlight the need to modify the expression of probability to preserve unitarity when the flavor states are nonorthogonal. 
We will continue to call this high scale unitarity violation in reference to a nonunitary $\mathbf{U}$. We contrast this to the low scale nonunitarity scenario in which there are new states accessible in neutrino oscillation experiments but the oscillations involving these states are fast enough such that they are averaged out. We further derive analytical formula for the neutrino oscillation amplitude involving $N$ neutrino flavors without assuming a unitarity $\mathbf{U}$ which allows us to prove a theorem that if $\left(\mathbf{U}\mathbf{U}^{\dagger}\right)_{\alpha\beta}=0$
for all $\alpha\neq\beta$, then the neutrino oscillation probability in an \emph{arbitrary matter potential} is \emph{indistinguishable} from the \emph{unitarity} scenario. 
Independently of matter potential, while nonunitarity effects for high scale nonunitarity scenario disappear as $\left(\mathbf{U}\mathbf{U}^{\dagger}\right)_{\alpha\beta}\to0$
for all $\alpha\neq\beta$, low scale nonunitarity effects can remain.

\end{abstract}

\maketitle
\flushbottom

\section{Introduction}

In the Standard Model (SM), there are three neutrinos which participate
in the weak interactions and we have detected all of them. Despite they should be massless in the SM, experimentally, we have determined
two nonzero mass-squared differences among them, showing that at least
two of them are massive. Great experimental progress has been made in pinning down the neutrino parameters in the three-flavor paradigm with the current global best fit values given by ~\cite{Esteban:2020cvm,NuFIT}: two mass splitting $\Delta m_{\textrm{sol}}^{2}\equiv m_{2}^{2}-m_{1}^{2}=7.4\times10^{-5}\,\textrm{eV}^{2}$, $\Delta m_{\textrm{atm}}^{2}\equiv \left|m_{3}^{2}-m_{1}^{2}\right|=2.5\times10^{-3}\,\textrm{eV}^{2}$, and three mixing angles
$\theta_{12}=33^\circ$, $\theta_{23}=49^\circ$, and $\theta_{13}=8.5^\circ$ determined to precision
of a few percents with a preference of Normal mass Ordering (NO) $m_3^2 - m_1^2 >0$. The absolute mass scale and the Dirac CP phase have not
been determined while $\theta_{23}$ can still be in the first or second quadrant.
From the theoretical side, the mechanism behind neutrino mass together
with the nature of the mass, Dirac or Majorana (including quasi-Dirac or pseudo-Dirac), remains an open question. 

Treating the SM as an effective field theory, Majorana mass for neutrinos
arise from the unique dimension-5 Weinberg operator \cite{Weinberg:1979sa,Weinberg:1980bf}. This
is the minimal scenario without additional light degrees of freedom.
In order to have Dirac mass, additional light degrees of freedom are
needed to be the Dirac partners of the SM neutrinos. In either cases, it
is not necessary that there is \emph{unitarity violation} or \emph{nonunitarity}
in the sense that the matrix $\mathbf{U}_{\alpha i}$ which relates
the flavor (with index $\alpha$) and mass (with index $i$) eigenstates
of neutrinos when sum over the kinematically accessible mass eigenstates or over the three SM flavors are not unitary
\begin{eqnarray}
\sum_{i}^{\textrm{accessible}}\mathbf{U}_{\alpha i}\mathbf{U}_{\beta i}^{*} & \neq & \delta_{\alpha\beta},\qquad\sum_{\alpha=e,\mu,\tau}\mathbf{U}_{\alpha i}\mathbf{U}_{\alpha j}^{*}\neq\delta_{ij}.\label{eq:nonunitarity}
\end{eqnarray}
In this work, we will focus on nonunitarity scenario when the relations above hold true and contrast it to the three-flavor paradigm when unitarity is preserved.

\begin{figure}
	\begin{centering}
		\includegraphics[scale=0.65]{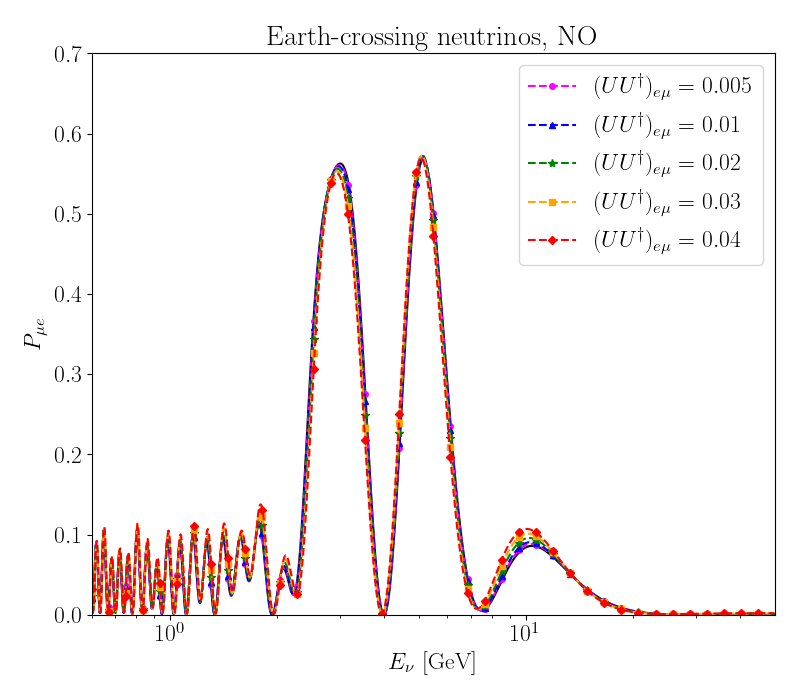}
		\par\end{centering}
	\caption{The probability of $\nu_\mu \to \nu_e$ for neutrino crossing the entire Earth as a function of neutrino energy $E_\nu$. We fix $\left(UU^{\dagger}\right)_{ee}=\left(UU^{\dagger}\right)_{\mu\mu}=0.96$, $\left(UU^{\dagger}\right)_{\tau\tau} = 1$ while allow $(UU^\dagger)_{e\mu}$ to vary.
	As the modulus of $(UU^\dagger)_{e\mu}$ decreases, the probability approaches that of the standard three-flavor unitarity scenario shown as the solid black line. \label{fig:high_scale_nonunitarity}}
\end{figure}

We aim to give a more complete theoretical discussion of nonunitarity in neutrino oscillation~\cite{Antusch:2006vwa,Fernandez-Martinez:2007iaa,Xing:2007zj,Goswami:2008mi,Antusch:2009pm,Xing:2011ur,Escrihuela:2015wra,Parke:2015goa,Dutta:2016vcc,Dutta:2016czj,Fong:2016yyh,Ge:2016xya,Blennow:2016jkn,Fong:2017gke,Martinez-Soler:2018lcy,Martinez-Soler:2019noy,Dutta:2019hmb,Ellis:2020hus,Wang:2021rsi,Forero:2021azc,Denton:2021mso,Agarwalla:2021owd,Majumdar:2022nby,Acero:2022wqg,Arguelles:2022tki}. 
We start by discussing how \emph{apparent} nonunitarity can arise in Section \ref{sec:models}, highlighting two distinct scenarios: \emph{high scale nonunitarity} scenario where new physics resides beyond the energy scale of neutrino oscillation experiments and \emph{low scale nonunitarity} scenario where new fermionic states (sterile neutrinos) mix with the SM neutrinos and are accessible in neutrino oscillation experiments.
In Section \ref{sec:theory}, we derive analytical solution for neutrino oscillation without assuming unitarity and in Section \ref{sec:high_vs_low}, we discuss how the high scale and low scale nonunitarity effects can show up and be distinguished in experiments. Finally we present some concluding remarks in Section \ref{sec:conclusions}. While many excellent discussions are there in previous work for e.g.~\cite{Antusch:2006vwa}, we will present some new results. In particular, we will prove a theorem in Section \ref{sec:theory} that if nonunitarity is only diagonal
\begin{eqnarray}
\sum_{i}^{\textrm{accessible}}\mathbf{U}_{\alpha i}\mathbf{U}_{\beta i}^{*} & = & c_{\alpha}\delta_{\alpha\beta},
\label{eq:diagonal_nonunitarity}
\end{eqnarray}
with $c_\alpha > 0$, then neutrino oscillation probability in an arbitrary matter potential is \emph{indistinguishable} from the \emph{unitarity} scenario. 
An important implication is that high scale nonunitarity effects are proportional to the off-diagonal elements of $\mathbf{U}\mathbf{U}^\dagger$, in contrast to low scale nonunitarity scenario where the effects can remain. 
To illustrate this point, in Figure \ref{fig:high_scale_nonunitarity}, we show the probability of $\nu_\mu \to \nu_e$ as a function of neutrino energy $E_\nu$ in high scale nonunitarity scenario for neutrino passing through the Earth using the public code \texttt{NuProbe}~\cite{Fong:2022oim,NuProbe} with a simplified (Preliminary Reference Earth Model) PREM model~\cite{Dziewonski:1981xy}. Denoting $U$ as a $3\times 3$ mixing matrix, we have fixed $(UU^{\dagger})_{ee}=(UU^{\dagger})_{\mu\mu}=0.96$, $(UU^{\dagger})_{\mu\mu} = 1$, $(UU^{\dagger})_{e\tau} = (UU^{\dagger})_{\mu\tau} = 0$ while varying $(UU^{\dagger})_{e\mu}$ from 0.005 to 0.04 (dashed lines) and the rest of the standard parameters have been fixed to the NO global best fit values from~\cite{Esteban:2020cvm,NuFIT}. As the modulus of $(UU^{\dagger})_{e\mu}$ decreases, the probability approaches that of the standard three-flavor unitarity scenario (solid black line). For reader interested in experimental probe of nonunitarity scenarios, he or she can jump straight to Section \ref{sec:high_vs_low}.

\section{Models}\label{sec:models}

Here we will focus on models for neutrino oscillation assuming that
the center-of-mass energy involved is below the electroweak symmetry
breaking $E<v_{\textrm{EW}}\equiv174$ GeV. We will consider only
Majorana mass term for neutrinos though the discussions below are independent
of whether we have a Majorana or Dirac mass term. While neutrino oscillation
cannot distinguish between strictly Majorana and Dirac mass, it is
possible to distinguish them from quasi-Dirac scenario in which both types of mass terms exist while the Majorana mass term is much smaller than the Dirac one~\cite{Cirelli:2004cz,deGouvea:2009fp,Anamiati:2017rxw,Anamiati:2019maf,Fong:2020smz}. We will consider this interesting scenario in a future
publication.

\subsection{High scale nonunitarity}
\label{subsec:high_scale_nonunitarity}

Assuming that for $E<v_{\textrm{EW}}$, we only have three SM neutrinos ($\nu_{e}$, $\nu_{\mu}$ and $\nu_{\tau}$ are
the SM left-handed neutrino flavor states), the general neutrino Lagrangian
allowed by the SM electromagnetic gauge symmetry $U(1)_{\textrm{EM}}$
in the charged lepton mass or flavor basis is given by
\begin{eqnarray}
{\cal L}_{\nu} & = & \frac{1}{2}\left(i\overline{\nu_{\alpha}}\slashed{\partial}D_{\alpha\beta}\nu_{\beta}-\overline{\nu_{\alpha}^{c}}m_{\alpha\beta}\nu_{\beta}+{\rm h.c.}\right)\nonumber \\
 &  & -\left(\frac{g}{2}W_{\mu}^{-}\overline{\ell_{\alpha}}\gamma^{\mu}P_{L}\nu_{\alpha}+\frac{g}{\sqrt{2}\cos\theta_{W}}Z_{\mu}\overline{\nu_{\alpha}}\gamma^{\mu}P_{L}\nu_{\alpha}+{\rm h.c.}\right),\label{eq:Lag_high_UV}
\end{eqnarray}
where $\alpha,\beta=e,\mu,\tau$ are flavor indices, $D$ is a
dimensionless Hermitian matrix while $m$ is a symmetric matrix with
mass dimension. In the second line, $g$ is the $SU(2)_{L}$ gauge
coupling, $\theta_{W}$ is the Weinberg or weak angle, $P_{L}=\frac{1}{\sqrt{2}}\left(1-\gamma^{5}\right)$
is the left-handed projector, $\left\{ \ell_{e},\ell_{\mu},\ell_{\tau}\right\} \equiv\left\{ e^{-},\mu^{-},\tau^{-}\right\} $
are the charged leptons and $W^{\mp}$ and $Z$ are the charged and
neutral weak bosons, respectively. Without additional light degrees of freedom,
only Majorana mass term is possible.\footnote{In order to write down a Dirac mass term, new light degrees of freedom
are required such that we can write $\overline{\nu'_{f}}m'_{f\beta}\nu_{\beta}$
where $\nu'_{f}$ are some new fermion fields which do not participate
in weak interactions. In this case, $m'_{f\beta}$ is a general complex
matrix with mass dimension. Besides the fact that the mass term should
be diagonalized by two unitary matrices, the discussion will remain
the same since the effect of unitary rotation of $\nu'$ which do
not feel the weak force, is not observable.} Treating the Standard Model as an effective field theory, the neutrino
mass and the modified kinetic terms come respectively from dimension-5 \cite{Weinberg:1979sa,Weinberg:1980bf}
and dimension-6 operators~\cite{Antusch:2006vwa,Broncano:2002rw,Antusch:2014woa,Fernandez-Martinez:2016lgt}
\begin{eqnarray}
{\cal O}_{5} & = & \frac{\lambda_{\alpha\beta}}{\Lambda_{5}}\left(\overline{L_{\alpha}^{c}}\epsilon H\right)\left(L_{\beta}^T\epsilon H\right),\\
{\cal O}_{6} & = & \frac{\eta_{\alpha\beta}}{\Lambda_{6}^{2}}\left(\overline{L_{\alpha}}\epsilon H^{*}\right)i\slashed{\partial}\left(L_{\beta}^T\epsilon H\right),
\end{eqnarray}
where $L_\alpha$ and $H$ are the $SU(2)_{L}$ lepton and Higgs doublets,
respectively, $\lambda$ and $\eta$ are dimensionless symmetric and
Hermitian matrices, respectively, and $\Lambda_{5}$ and $\Lambda_{6}$
are effective scales below which the operators ${\cal O}_{5}$ and
${\cal O}_{6}$ are valid. Implicitly, we have assumed $E\ll\Lambda_{5},\Lambda_{6}$.
Not all ultraviolet models which generate ${\cal O}_{5}$ also generate
${\cal O}_{6}$. For instance, type-I and type-III seesaw models generate
both ${\cal O}_{5}$ and ${\cal O}_{6}$ while type-II seesaw model
only generates ${\cal O}_{5}$.

In order to obtain canonical normalized kinetic term, we can first
diagonalize the kinetic term as $D=Y^{\dagger}\hat{D}Y$ where $Y$
is unitary and $\hat{D}$ is real and diagonal. Defining the normalized
neutrino fields as $\widetilde{\nu}\equiv\sqrt{\hat{D}}Y\nu$, eq.
(\ref{eq:Lag_high_UV}) becomes
\begin{eqnarray}
{\cal L}_{\nu} & = & \frac{1}{2}\left(i\overline{\widetilde{\nu}_{\alpha}}\slashed{\partial}\widetilde{\nu}_{\alpha}-\overline{\widetilde{\nu}_{\alpha}^{c}}\widetilde{m}_{\alpha\beta}\widetilde{\nu}_{\beta}+{\rm h.c.}\right)\nonumber \\
 &  & -\left[\frac{g}{2}W_{\mu}^{-}\overline{\ell_{\alpha}}\gamma^{\mu}P_{L}\left(Y^{\dagger}\sqrt{\hat{D}}^{-1}\right)_{\alpha\beta}\widetilde{\nu}_{\beta}+\frac{g}{\sqrt{2}\cos\theta_{W}}Z_{\mu}\left(\hat{D}^{-1}\right)_{\alpha\alpha}\overline{\widetilde{\nu}_{\alpha}}\gamma^{\mu}P_{L}\widetilde{\nu}_{\alpha}+{\rm h.c.}\right],
\end{eqnarray}
where we have defined
\begin{eqnarray}
\widetilde{m} & \equiv & \sqrt{\hat{D}}^{-1}Y^{*}mY^{\dagger}\sqrt{\hat{D}}^{-1}.
\end{eqnarray}
The symmetric mass matrix above can be diagonalized by a unitary matrix
$V$ as $\widetilde{m}=V^{*}\hat{m}V^{\dagger}$ where $\hat{m}$
is real and diagonal. Defining the neutrino fields in the mass basis
as $\hat{\nu}=V^{\dagger}\widetilde{\nu}$, we have~\cite{Antusch:2006vwa}
\begin{eqnarray}
{\cal L}_{\nu} & = & \frac{1}{2}\left(i\overline{\hat{\nu}_{i}}\slashed{\partial}\hat{\nu}_{i}-\overline{\hat{\nu}_{i}^{c}}\hat{m}_{ii}\hat{\nu}_{i}+{\rm h.c.}\right)\nonumber \\
 &  & -\left[\frac{g}{2}W_{\mu}\overline{\ell_{\alpha}}\gamma^{\mu}P_{L}U_{\alpha i}\hat{\nu}_{i}+\frac{g}{\sqrt{2}\cos\theta_{W}}Z_{\mu}\overline{\hat{\nu}_{i}}\left(U^{\dagger}U\right)_{ij}\gamma^{\mu}P_{L}\hat{\nu}_{j}+{\rm h.c.}\right],
\end{eqnarray}
where we denote $i,j=1,2,3$ to be the indices in mass basis and we
have defined
\begin{eqnarray}
U_{\alpha i} & \equiv & \left(Y^{\dagger}\sqrt{\hat{D}}^{-1}V\right)_{\alpha i}.
\end{eqnarray}
Notice that
\begin{eqnarray}
UU^{\dagger} & = & Y^{\dagger}\hat{D}^{-1}Y,\qquad U^{\dagger}U=V^{\dagger}\hat{D}^{-1}V.
\end{eqnarray}
Only if $\hat{D}=I$ is the $3\times3$ identity matrix, unitarity is
restored in which $U^{\dagger}U=UU^{\dagger}=I$. We denote the general
case with $UU^{\dagger}\neq I$ and $U^{\dagger}U\neq I$, \emph{high
scale} nonunitarity scenario. 
Later, we will prove that scenario with $(UU^{\dagger})_{\alpha\beta} = 0$ for all $\beta\neq\alpha$ is \emph{indistinguishable} from unitarity scenario even if $(UU^{\dagger})_{\alpha\alpha} \neq 1$.

\subsection{Low scale nonunitarity}
\label{subsec:low_scale_nonunitarity}

Assuming that for $E<v_{\textrm{EW}}$, besides the three SM neutrinos
($\nu_{e}$, $\nu_{\mu}$ and $\nu_{\tau}$), we also have additional
neutral fermions fields ($\nu_{s_1}$, $\nu_{s_2}$,..., $\nu_{s_N}$) which
do not participate in weak interactions but mix with the SM neutrinos
through the mass term. The mixing between the SM and the additional
fermions are the Dirac mass term. Here we will focus on mostly Majorana
scenario where the Majorana mass term for new fermions is somewhat
larger than the Majorana mass for the SM neutrinos as well as the
Dirac mass term.\footnote{The situation where the Dirac mass term is much larger than the Majorana mass term results in quasi-Dirac or pseudo-Dirac scenario with distinguished signatures~\cite{Cirelli:2004cz,deGouvea:2009fp,Anamiati:2017rxw,Anamiati:2019maf,Fong:2020smz} and will be considered in a future work.}
 
In order to highlight the distinction from the high scale unitarity
violation scenario, we further assume that the kinetic terms for all
the fermions are canonical. In this case, the $U(1)_{\textrm{EM}}$-invariant
neutrino Lagrangian in the charged lepton flavor basis is given by
\begin{eqnarray}
{\cal L}_{\nu} & = & \frac{1}{2}\left(i\overline{\nu_{\tilde\alpha}}\slashed{\partial}\nu_{\tilde\alpha}-\overline{\nu_{\tilde\alpha}^{c}}m_{\tilde\alpha\tilde \beta}\nu_{\tilde\beta}+{\rm h.c.}\right)\nonumber \\
 &  & -\left(\frac{g}{2}W_{\mu}^{-}\overline{\ell_{\alpha}}\gamma^{\mu}P_{L}\nu_{\alpha}+\frac{g}{\sqrt{2}\cos\theta_{W}}Z_{\mu}\overline{\nu_{\alpha}}\gamma^{\mu}P_{L}\nu_{\alpha}+{\rm h.c.}\right),
\end{eqnarray}
where $\alpha,\beta,...=e,\mu,\tau$ 
while $\tilde\alpha,\tilde\beta,...=e,\mu,\tau,s_{1},s_{2},...,s_{N}$. The symmetric
mass matrix $m$ can be diagonalized by a unitary matrix $\mathbf{U}$
as $m=\mathbf{U}^{*}\hat{m}\mathbf{U}^{\dagger}$ where $\hat{m}$
is real and diagonal. Here and in the following, we use a boldface
$\mathbf{U}$ to denote a $\left(3+N\right)\times\left(3+N\right)$
matrix while $U$ is reserved for a $3\times3$ matrix. Defining the
neutrino field in the mass basis as $\hat{\nu}=\mathbf{U}^{\dagger}\nu$,
we have
\begin{eqnarray}
{\cal L}_{\nu} & = & \frac{1}{2}\left(i\overline{\hat{\nu}_{i}}\slashed{\partial}\hat{\nu}_{i}-\overline{\hat{\nu}_{i}^{c}}\hat{m}_{ii}\hat{\nu}_{i}+{\rm h.c.}\right)\nonumber \\
 &  & -\left[\frac{g}{2}W_{\mu}\overline{\ell_{\alpha}}\gamma^{\mu}P_{L}\mathbf{U}_{\alpha i}\hat{\nu}_{i}+\frac{g}{\sqrt{2}\cos\theta_{W}}Z_{\mu}\overline{\hat{\nu}_{i}}\mathbf{U}_{\alpha i}^*
 \gamma^{\mu}P_{L}\mathbf{U}_{\alpha j}\hat{\nu}_{j}+{\rm h.c.}\right],
\end{eqnarray}
where $i,j,...=1,2,...,3+N$. What distinguish this from the high
scale nonunitarity scenario is that the flavor states remain \emph{orthogonal} since $\mathbf{U}$ is unitary $\mathbf{U}^{\dagger}\mathbf{U}=\mathbf{U}\mathbf{U}^{\dagger}=\mathbf{I}$
where $\mathbf{I}$ is the $\left(3+N\right)\times\left(3+N\right)$
identity matrix. 

Strictly speaking, there is no unitarity violation in this case. However,
since only the mixing involving the SM neutrinos can be measured,
one have
\begin{eqnarray}
\sum_{\alpha=e,\mu,\tau}\mathbf{U}_{\alpha i}\mathbf{U}_{\alpha j}^{*} & \neq & \delta_{ij}.
\end{eqnarray}
Furthermore, if $m_{i>3}\gg \sqrt{\Delta m_\textrm{atm}^2}$ such that oscillations
involving $\nu_{i>3}$ can be averaged out, at leading order in small
unitarity violating parameter, the mixing elements involved are those
of $i=1,2,3$ which sum to~\cite{Fong:2016yyh,Fong:2017gke} 
\begin{eqnarray}
\sum_{i=1}^{3}\mathbf{U}_{\alpha i}\mathbf{U}_{\beta i}^{*} & \neq & \delta_{\alpha\beta}.
\end{eqnarray}
We denote this as \emph{low scale} nonunitarity scenario. If some additional
states are not kinematically allowed in the process, one should describe
it as in the high scale nonunitarity scenario discussed previously,
with possible enlargement of flavor space beyond three dimensions to accommodate additional kinematically accessible states.

\section{Generic $3+N$ neutrino oscillations}\label{sec:theory}

Here we will develop a generic $3+N$ neutrino oscillation framework
which can be applied to neutrino oscillation with either unitary or nonunitarity $\mathbf{U}$.
In general, the neutrino flavor states $\left|\nu_{\alpha}\right\rangle $
are related to the mass eigenstates $\left|\nu_{i}\right\rangle $
through a matrix $\mathbf{U}$
\begin{eqnarray}
\left|\nu_{\alpha}\right\rangle  & = & \frac{1}{\sqrt{\left(\mathbf{U}\mathbf{U}^{\dagger}\right)_{\alpha\alpha}}}\sum_{i}\mathbf{U}_{\alpha i}^{*}\left|\nu_{i}\right\rangle ,\label{eq:mass_to_flavor}
\end{eqnarray}
where $\alpha,\beta,...=e,\mu,\tau,s_{1},s_{2},...,s_{N}$ ($\nu_{e}$,
$\nu_{\mu}$ and $\nu_{\tau}$ are the SM left-handed neutrino flavor
states) and $i,j,...=1,2,...,3+N$. In general, $\mathbf{U}$ does
not have to be unitary $\mathbf{U}\mathbf{U}^{\dagger}\neq\mathbf{I}$
and $\mathbf{U}^{\dagger}\mathbf{U}\neq\mathbf{I}$. Since the mass
eigenstates are orthogonal $\left\langle \nu_{j}|\nu_{i}\right\rangle =\delta_{ji}$,
the flavor states are properly normalized $\left\langle \nu_{\alpha}|\nu_{\alpha}\right\rangle =1$
though they are not necessarily orthogonal but equal to
\begin{eqnarray}
\left\langle \nu_{\beta}|\nu_{\alpha}\right\rangle  & = & \frac{\left(\mathbf{U}\mathbf{U}^{\dagger}\right)_{\beta\alpha}}{\sqrt{\left(\mathbf{U}\mathbf{U}^{\dagger}\right)_{\alpha\alpha}\left(\mathbf{U}\mathbf{U}^{\dagger}\right)_{\beta\beta}}}.\label{eq:nonorthogonality}
\end{eqnarray}
In other words, if there is nonzero overlap between different flavor states and there is a probability of ``\emph{flavor changing}'' even at zero distance.\footnote{More precisely, given a state $\nu_{\alpha}$, there is a probability of measuring it as $\nu_{\beta\neq\alpha}$ since the two states are not orthogonal.}
In the rest of the article, to avoid expressions crowded with normalization
factors, we will define
\begin{eqnarray}
	\overline{\mathbf{U}}_{\alpha i} & \equiv & \frac{\mathbf{U}_{\alpha i}}{\sqrt{\left(\mathbf{U}\mathbf{U}^{\dagger}\right)_{\alpha\alpha}}}.
\end{eqnarray}

From eq. (\ref{eq:mass_to_flavor}), we can write the inverse relation\footnote{We can prove that inverse exists. Supposing that 
	\begin{eqnarray*}
		\left|\nu_{i}\right\rangle  & = & \sum_{\alpha}\mathbf{V}_{i\alpha}\left|\nu_{\alpha}\right\rangle ,
	\end{eqnarray*}
	and using orthogonality condition $\left\langle \nu_{j}|\nu_{i}\right\rangle =\delta_{ji}$,
	we have
	\begin{eqnarray*}
		\delta_{ij} & = & \sum_{\alpha}\mathbf{V}_{i\alpha}\left\langle \nu_{j}|\nu_{\alpha}\right\rangle
		=\sum_{\alpha}\mathbf{V}_{i\alpha}\overline{\mathbf{U}}_{\alpha j}^{*}, 
	\end{eqnarray*}
	and hence
	$\mathbf{V}_{i\alpha}=\left(\overline{\mathbf{U}}^{*,-1}\right)_{i\alpha}$.
}
\begin{eqnarray}
	\left|\nu_{i}\right\rangle  & = &
	\sum_{\alpha}\left(\overline{\mathbf{U}}^{*,-1}\right)_{i\alpha}\left|\nu_{\alpha}\right\rangle, 
	\label{eq:flavor_to_mass}
\end{eqnarray}
which as a check, verifies the orthogonality of mass eigenstates
\begin{eqnarray}
	\left\langle \nu_{j}|\nu_{i}\right\rangle  & = &
	\sum_{\alpha,\beta}\left(\overline{\mathbf{U}}^{-1}\right)_{j\beta}
	\left(\overline{\mathbf{U}}^{*,-1}\right)_{i\alpha}\left\langle \nu_{\beta}|\nu_{\alpha}\right\rangle 
	=\left(\overline{\mathbf{U}}^{-1}\overline{\mathbf{U}}\,\overline{\mathbf{U}}^{\dagger}\overline{\mathbf{U}}^{\dagger,-1}\right)_{ji}=\delta_{ji},
\end{eqnarray}
where in the second equality, we have used eq. (\ref{eq:nonorthogonality}).

\subsection{Completeness and unitarity}

The set of all $\left\{ \left|\nu_{i}\right\rangle \right\} $
is complete and from the orthogonality condition, it further satisfies
the completeness relation
\begin{eqnarray}
\sum_{i}\left|\nu_{i}\right\rangle \left\langle \nu_{i}\right| & = & \mathbf{I}.\label{eq:completeness}
\end{eqnarray}
Inserting the completeness relation above in between $\langle \nu_\alpha|\nu_\alpha\rangle=1$ implies that the probabilities of a flavor state $\left|\nu_{\alpha}\right\rangle$ being detected as all possible mass eigenstates sum up to unity $\displaystyle\sum_{i} \left|\left\langle \nu_i | \nu_\alpha\right\rangle\right|^2 = 1$ as required. 
	
For general flavor basis set $\left\{ \left|\nu_{\alpha}\right\rangle \right\}$ which is complete but not necessarily orthogonal, it satisfies a modified completeness relation\footnote{See Appendix \ref{app:nonorthogonal} for derivation.}
\begin{eqnarray}
	\sum_{\alpha,\beta}  \left|\nu_\alpha\right\rangle \left[\big(\overline{\mathbf{U}}\, \overline{\mathbf{U}}^\dagger\big)^{-1} \right]_{\alpha\beta} \left\langle \nu_\beta \right| &=& \mathbf{I},
	\label{eq:completeness_flavor}
\end{eqnarray}	
taking into account the possible overlaps between the flavor states. Inserting the relation above into $\langle \nu_i|\nu_i \rangle = 1$, we obtain\footnote{This relation can also be verified explicitly using eq.~\eqref{eq:mass_to_flavor}.}
\begin{eqnarray}
	\sum_\alpha \left|\left\langle \nu_\alpha| \nu_i\right\rangle \right|^2 
	+ \sum_{\alpha}\sum_{\beta \neq \alpha}  
	\left\langle \nu_i| \nu_\alpha\right\rangle
	\left[\big(\overline{\mathbf U} \,\overline{\mathbf U}^\dagger\big)^{-1}\right]_{\alpha\beta} \left\langle \nu_\beta| \nu_i\right\rangle
	&=& 1.
	\label{eq:probability_summation_nonorthogoanl}
\end{eqnarray}
For nonorthogonal flavor states, we can no longer interpret the first term as the sum of the probabilities of a mass eigenstate $\nu_i$ to be measured in all possible flavor eigenstates $\nu_\alpha$. In this case, the \emph{correct} probability of detecting a flavor state $\nu_\alpha$ from $\nu_i$ has to include the contributions from other flavor states as follows
\begin{eqnarray}
	P_{\alpha i} = \left|\left\langle \nu_\alpha| \nu_i\right\rangle \right|^2 +
	\sum_{\beta \neq \alpha}  
	\left\langle \nu_i| \nu_\alpha\right\rangle
	\left[\big(\overline{\mathbf U} \,\overline{\mathbf U}^\dagger\big)^{-1}\right]_{\alpha\beta} \left\langle \nu_\beta| \nu_i\right\rangle,
\end{eqnarray}
and from eq.~\eqref{eq:probability_summation_nonorthogoanl}, summing over $\alpha$ gives unity as required. With orthogonal flavor states $\left[\big(\overline{\mathbf U} \,\overline{\mathbf U}^\dagger\big)^{-1}\right]_{\alpha\beta} = \delta_{\alpha\beta}$, one recover the standard result.

Since the flavor state space is complete, one also expect the oscillation probability of $\nu_{\alpha}$ to all possible final flavor states to sum up to one
\begin{eqnarray}
	\sum_\beta P_{\beta\alpha} = 1.
	\label{eq:flavor_unitarity}
\end{eqnarray}
However, considering possible nonorthogonal flavor states, the probability will be modified from the usual expression  $P_{\beta\alpha} = |\left\langle \nu_{\beta}|\nu_{\alpha}\right\rangle|^2$. In Section \ref{sec:osc_prob_nonorthogonal}, we will discuss the probability operator which gives rise to probability that preserves unitarity as in eq.~\eqref{eq:flavor_unitarity}.

\subsection{Evolution of a flavor state}\label{subsec:evolution_flavor_state}

Starting from an initial state $\left|\nu_{\alpha}\left(0\right)\right\rangle =\left|\nu_{\alpha}\right\rangle $,
the time-evolved state $\left|\nu_{\alpha}\left(t\right)\right\rangle $
is described by the Schrödinger equation
\begin{eqnarray}
i\frac{d}{dt}\left|\nu_{\alpha}\left(t\right)\right\rangle  & = & {\cal H}\left|\nu_{\alpha}\left(t\right)\right\rangle ,\label{eq:Schr_eq}
\end{eqnarray}
where the Hamiltonian is ${\cal H}={\cal H}_{0}+{\cal H}_{I}$ with
${\cal H}_{0}$ the free Hamiltonian 
\begin{eqnarray}
{\cal H}_{0}\left|\nu_{i}\right\rangle  & = & E_{i}\left|\nu_{i}\right\rangle ,\;\;\;\;\;E_{i}=\sqrt{\vec{p}_{i}^{\;~2}+m_{i}^{2}},\label{eq:H0}
\end{eqnarray}
and ${\cal H}_{I}$ the interaction Hamiltonian with matrix elements
\begin{eqnarray}
\left\langle \nu_{\beta}\right|{\cal H}_{I}\left|\nu_{\alpha}\right\rangle  & = & V_{\beta\alpha}.\label{eq:HI}
\end{eqnarray}
Since ${\cal H}_{I}^{\dagger}={\cal H}_{I}$, we have $V_{\beta\alpha}^{*}=V_{\alpha\beta}$.

Assuming relativistic neutrinos, we trade $t=x$ and the amplitude
of the transition $\nu_{\alpha}\to\nu_{\beta}$ at distance $x$ is
then given by $S_{\beta\alpha}\left(x\right)\equiv\left\langle \nu_{\beta}|\nu_{\alpha}\left(x\right)\right\rangle$.
From eq.~\eqref{eq:Schr_eq}, we can write the evolution equation of $S_{\beta\alpha}\left(x\right)$
as
\begin{eqnarray}
i\frac{d}{dx}S_{\beta\alpha}\left(x\right) & = & \left\langle \nu_{\beta}\right|{\cal H}_{0}+{\cal H}_{I}\left|\nu_{\alpha}\left(t\right)\right\rangle \nonumber \\
 & = & \sum_{i}\left\langle \nu_{\beta}\right|{\cal H}_{0}+{\cal H}_{I}\left|\nu_{i}\right\rangle \left\langle \nu_{i}|\nu_{\alpha}\left(t\right)\right\rangle \nonumber \\
 & = & \sum_{\eta}\left\{\sum_{i}\overline{\mathbf{U}}_{\beta i}E_{i}
 \big(\overline{\mathbf{U}}^{-1}\big)_{i\eta}
 +\sum_{\gamma}V_{\beta\gamma}\left[\big(\overline{\mathbf{U}}\,\overline{\mathbf{U}}^{\dagger}\big)^{-1}\right]_{\gamma\eta}\right\}S_{\eta\alpha}\left(x\right),
\end{eqnarray}
where in the second equality, we have inserted the completeness relation
eq. (\ref{eq:completeness}) and in the last equality, we have used
eqs. (\ref{eq:H0}), (\ref{eq:HI}), (\ref{eq:flavor_to_mass}) and
(\ref{eq:nonorthogonality}). Considering relativistic neutrinos $E\gg m_{i}$
and expanding $E_{i}\simeq E+\frac{m_{i}^{2}}{2E}$, we obtain, in
matrix notation
\begin{eqnarray}
i\frac{dS\left(x\right)}{dx} & = & \left[\overline{\mathbf{U}}\Delta\overline{\mathbf{U}}^{-1}
+V\big(\overline{\mathbf{U}}\,\overline{\mathbf{U}}^{\dagger}\big)^{-1}\right]S\left(x\right),\label{eq:evol_flavor_basis}
\end{eqnarray}
where
\begin{eqnarray}
\Delta & \equiv & \frac{1}{2E}\textrm{diag}\left(m_{1}^{2},m_{2}^{2},...,m_{3+N}^{2}\right)=\textrm{diag}\left(\Delta_{1},\Delta_{2},...,\Delta_{3+N}\right).\label{eq:Delta}
\end{eqnarray}
We have dropped the constant $E$ which is an overall phase in $S(x)$ and not observable.

\subsection{Vacuum mass basis}

From eq. (\ref{eq:evol_flavor_basis}), the Hamiltonian in the flavor basis given by
\begin{eqnarray}
H & \equiv & \overline{\mathbf{U}}\Delta\overline{\mathbf{U}}^{-1}
+V\big(\overline{\mathbf{U}}\,\overline{\mathbf{U}}^{\dagger}\big)^{-1},\label{eq:H_flavor_basis}
\end{eqnarray}
is not Hermitian $H^{\dagger}\neq H$ nor \emph{normal} $H^{\dagger}H\neq HH^{\dagger}$.
In the following, we will prove that they can be diagonalized with real eigenvalues. Furthermore, we will argue that despite the apparent non-Hermitian Hamiltonian, unitarity is actually preserved.

Let us change the Hamiltonian to the \emph{vacuum mass basis} in which
the free Hamiltonian is diagonal
\begin{eqnarray}
\widetilde{H} & \equiv & \overline{\mathbf{U}}^{-1}H\overline{\mathbf{U}}=\Delta+\overline{\mathbf{U}}^{-1}V\overline{\mathbf{U}}^{\dagger,-1}.\label{eq:H_mass_basis}
\end{eqnarray}
Notice that eq. (\ref{eq:H_mass_basis}) is Hermitian $\widetilde{H}^{\dagger}=\widetilde{H}$.
Assuming $V=V^{\dagger}$ is constant in the interval of interest
$0\leq x<x_{1}$, we can diagonalize the $\widetilde{H}$ with a unitary
matrix $X$
\begin{eqnarray}
\widetilde{H} & = & X\hat{H}X^{\dagger},\label{eq:X_diagonalization}
\end{eqnarray}
where $X$ is unitary and $\hat{H}=\textrm{diag}\left(\lambda_{1},\lambda_{2},...,\lambda_{3+N}\right)$ is diagonal and real. 
Since eqs. (\ref{eq:H_flavor_basis}) and (\ref{eq:H_mass_basis})
are related by similar transformation, they have the same eigenvalues
$\widetilde{H}X=X\hat{H}\implies H\overline{\mathbf{U}}X=\overline{\mathbf{U}}X\hat{H}$ and hence we can also write
\begin{eqnarray}
H & = & \overline{\mathbf{U}}X\hat{H}\left(\overline{\mathbf{U}}X\right)^{-1},\label{eq:diagonalization_nonHermitian}
\end{eqnarray}
where the nonnormal $H$ is diagonalized by a nonunitary $\overline{\mathbf{U}}X$.
We see explicitly that despite the Hamiltonian in flavor basis $H$ appears
to be non-Hermitian, the eigenvalues remain real while the source
of non Hermicity comes from nonunitary transformation matrix $\mathbf{U}$. Although $\overline{\mathbf{U}}X$ is nonunitary, one can formally solve for $\left(\overline{\mathbf{U}}X\right)_{\alpha i}\left(\overline{\mathbf{U}}X\right)_{i\beta}^{-1}$ in terms of eigenvalues and Hamiltonian elements using the same method as in refs.~\cite{Yasuda:2007jp,Fong:2022oim}. 
Nevertheless, as we will see in the next subsection, 
the combination which appears in neutrino oscillation
probability is not $\left(\overline{\mathbf{U}}X\right)_{\alpha i}\left(\overline{\mathbf{U}}X\right)_{i\beta}^{-1}$ but $\left(\overline{\mathbf{U}}X\right)_{\alpha i}\left(\overline{\mathbf{U}}X\right)_{i\beta}^\dagger$ and hence we will solve for $X_{ik} X_{jk}^*$ instead.

Now let us pause to ask a valid question: do we expect unitarity to be violated? In the vacuum mass basis, since the Hamiltonian \eqref{eq:H_mass_basis} is Hermitian, unitarity should be preserved under time evolution. By doing a similarity transformation with non-Hermitian $\overline{\mathbf{U}}$ back to the flavor basis Hamitonian \eqref{eq:H_flavor_basis} appears to be non-Hermitian but this is just an apparent feature. As long as a non-Hermitian Hamiltonian can be transformed to a Hermitian Hamiltonian by similarity transformation, unitarity is preserved. This is also expected from the outset since the flavor space is complete.

As shown in refs.~\cite{Yasuda:2007jp,Fong:2022oim}, by raising eq. (\ref{eq:X_diagonalization})
to the power of $1,2,...,2+N$ and taking into account the unitarity
relation $XX^{\dagger}=\mathbf{I}$, one can form a set of $3+N$ linearly
independent equations for $X_{ik}X_{jk}^{*}$ where the coefficients
form a Vandermonde matrix\footnote{The equations obtained with power in $\lambda_{i}$ greater than $2+N$
	are not linearly independent since they can be rewritten in term of
	lower power using the characteristic equation of $\widetilde{H}$.
	Suppose we have $d+1$ degenerate eigenvalues $\lambda_{l}=\lambda_{k}$
	for $l=k,...,k+d$, we only need to solve for the combination $\sum_{l}X_{il}X_{jl}^{*}$
	corresponding to $\lambda_{k}$, i.e. $3+N-d$ linear equations
    can be obtained from raising eq. (\ref{eq:X_diagonalization})
	to the power of $1,2,...,2+N-d$, including $XX^{\dagger}=\mathbf{I}$.} which can be inverted to give~\cite{Fong:2022oim} (see also the pioneering work of Kimura, Takamura and Yokomakura who applied similar method for 3-flavor scenario~\cite{Kimura:2002hb,Kimura:2002wd})
\begin{eqnarray}
X_{ik}X_{jk}^{*} & = & \frac{{\displaystyle \sum_{p=0}^{2+N}}\left(-1\right)^{p}(\widetilde{H}^{p})_{ij}c_{2+N-p,k}}{Z_{k}},\label{eq:main_result}
\end{eqnarray}
where we have defined
\begin{eqnarray}
Z_{k} & \equiv & \prod_{p\neq k}\left(\lambda_{p}-\lambda_{k}\right),\\
c_{p,k} & \equiv & \sum_{\left\{ q\neq r\neq...\right\} \neq k}\underbrace{\lambda_{q}\lambda_{r}...}_{p},
\end{eqnarray}
with $[\tilde{H}^{0}]_{ij}=\delta_{ij}$ and $c_{0,k}=1$.
The sum in $c_{p,k}$ is over all possible unordered combinations
of $p$ distinct eigenvalues $\lambda_{q}\lambda_{r}...$ where none
of them is equal to $\lambda_{k}$ and hence with $3+N$ neutrino flavors,
$c_{p,k}$ has $\left(\begin{smallmatrix}2+N \\ p
\end{smallmatrix}\right) = \frac{\left(2+N\right)!}{p!\left(2+N-p\right)!}$  terms in the sum. 
As shown in ref.~\cite{Abdullahi:2022fkh}, the numerator of eq. (\ref{eq:main_result}) can be written in mathematically
equivalent form in terms of elements of the adjugate of $\lambda_{k}I-\widetilde{H}$
i.e. $[\textrm{Adj}(\lambda_{k}I-\widetilde{H})]_{ij}$ and can be equally fast in numerical evaluation using
the Le Verrier-Faddeev algorithm. 

\subsection{Oscillation probability}

There is one subtle but \emph{crucial} point regarding $\widetilde{S}$
in the vacuum mass basis which satisfies
\begin{eqnarray}
i\frac{d\widetilde{S}\left(x\right)}{dx} & = & \widetilde{H}\widetilde{S}\left(x\right),\label{eq:evol_mass_basis}
\end{eqnarray}
and the solution is given by
\begin{eqnarray}
\widetilde{S}\left(x\right) & = & Xe^{-i\hat{H}x}X^{\dagger}.
\end{eqnarray}
How do we relate $\widetilde{S}\left(x\right)$ of the vacuum mass basis and $S\left(x\right)$
in the flavor basis? Fixing the initial conditions $\widetilde{S}\left(0\right)=\mathbf{I}$
and $S\left(0\right)=\overline{\mathbf{U}}\,\overline{\mathbf{U}}^{\dagger}$
which follow from the orthogonality of mass eigenstates and the nonorthogonality
of flavor eigenstates eq. (\ref{eq:nonorthogonality}), respectively,
we have
\begin{eqnarray}
\widetilde{S}\left(x\right) & \equiv & \overline{\mathbf{U}}^{-1}S\left(x\right)\overline{\mathbf{U}}^{\dagger,-1}.\label{eq:Stilde}
\end{eqnarray}
Hence we can write\footnote{For antineutrino $\overline{\nu}_{\alpha}\to\overline{\nu}_{\beta}$, we take $\overline{\mathbf{U}}_{\alpha i}\to\overline{\mathbf{U}}_{\alpha i}^{*}$ and since our Universe consists only of matter, we should also take $V\to-V$ in eq. (\ref{eq:evol_mass_basis}).}
\begin{eqnarray}
S_{\beta\alpha}\left(x\right) & = & \left[\overline{\mathbf{U}}Xe^{-i\hat{H}x}\left(\overline{\mathbf{U}}X\right)^{\dagger}\right]_{\beta\alpha}
=\sum_{i,j,k}\overline{\mathbf{U}}_{\beta i}\overline{\mathbf{U}}_{\alpha j}^{*}X_{ik}X_{jk}^{*}e^{-i\lambda_{k}x},
\label{eq:S_constant_V}
\end{eqnarray}
which has exactly the same form as the unitarity case despite
that $\overline{\mathbf{U}}$ does not have to be unitary. 

The probability of an initial state $\left|\nu_{\alpha}\right\rangle $ being detected as $\left|\nu_{\beta}\right\rangle $ at distance $0<x<x_1$ (where $V(x)$ is constant) is
\begin{eqnarray}
	P_{\beta\alpha}\left(x\right) & = & \sum_{\xi,\lambda}
	S_{\alpha\xi}(x) (\hat P_{\beta})_{\xi\lambda} S_{\lambda\alpha}(x).
	\label{eq:prob}
\end{eqnarray}
The appearance of $(\hat P_{\beta})_{\xi\lambda}$ takes into account possible nonorthogonality of flavor states.
Since this is a nontrivial result with subtleties, the discussion of $(\hat P_{\beta})_{\xi\lambda}$ will be deferred to Section~\ref{sec:osc_prob_nonorthogonal}. 
For the moment, note that summing over $\beta$, we have
\begin{eqnarray}
	\sum_{\beta}(\hat P_{\beta})_{\xi\lambda} = \left[\big(\overline{\mathbf{U}}\,\overline{\mathbf{U}}^{\dagger}\big)^{-1}\right]_{\xi\lambda}.
	\label{eq:unitarity_preservation}
\end{eqnarray}
Together with eq.~\eqref{eq:completeness_flavor}, this guarantees that $\sum_{\beta} P_{\beta\alpha}(x) = 1$. For orthogonal flavor states, we have
\begin{eqnarray}
(\hat P_{\beta})_{\xi\lambda} = \delta_{\xi\beta}\delta_{\lambda\beta},
 	\label{eq:orthogonal_op}
\end{eqnarray}
which gives the standard expression
$P_{\beta\alpha}\left(x\right) = \left|S_{\beta\alpha}\left(x\right)\right|^{2}$.
Using eq.~\eqref{eq:orthogonal_op} for nonorthogonal flavor states will lead inconsistent result which violates unitarity $\sum_\beta P_{\alpha\beta} \neq 1$. By substituting eq.~\eqref{eq:orthogonal_op} into eq.~\eqref{eq:prob}, it is sufficient to show this violation at zero distance $x=0$ with $X_{ij} = \delta_{ij}$ in which we obtain
\begin{eqnarray}
P_{\beta\alpha}\left(0\right) & = &
(\overline{\mathbf{U}}\,\overline{\mathbf{U}}^\dagger)_{\alpha\beta}
(\overline{\mathbf{U}}\,\overline{\mathbf{U}}^\dagger)_{\beta\alpha} = |(\overline{\mathbf{U}}\,\overline{\mathbf{U}}^\dagger)_{\beta\alpha}|^2,
\end{eqnarray}
which implies $\sum_{\beta} P_{\beta\alpha}(0) > 1$ if the flavor states are nonorthogonal since $(\overline{\mathbf{U}}\,\overline{\mathbf{U}}^\dagger)_{\alpha\alpha} = 1$.

One can generalize the solution in eq. (\ref{eq:S_constant_V}) to
the case when $V$ is $x$-dependent by splitting $x$ into intervals
small enough that $V\left(x\right)$ is approximately constant. Considering
$0=x_{0}<x_{1}<x_{2}<...$ where $V\left(x\right)$ is equal to constant
$V_{a}$ for each interval $x_{a-1}<x<x_{a}$, we obtain
\begin{eqnarray}
S\left(x\right) & = & T\prod_{a=1}
\big(\overline{\mathbf{U}}\,\overline{\mathbf{U}}^{\dagger}\big)^{-1}
S^{\left(a\right)}\left(x\right),\label{eq:S_x_dependent_potential}
\end{eqnarray}
where we have defined
\begin{eqnarray}
S^{\left(a\right)}\left(x\right) & \equiv & \left(\overline{\mathbf{U}}X^{\left(a\right)}\right)e^{-i\hat{H}^{\left(a\right)}x^{\left(a\right)}}\left(\overline{\mathbf{U}}X^{\left(a\right)}\right)^{\dagger},\label{eq:S_a}\\
x^{\left(a\right)} & \equiv & \left[\left(x-x_{a-1}\right)\theta\left(x_{a}-x\right)+\left(x_{a}-x_{a-1}\right)\theta\left(x-x_{a}\right)\right]\theta\left(x-x_{a-1}\right),\label{eq:x_a}
\end{eqnarray}
with $\theta\left(x\geq0\right)=1$, $\theta\left(x<0\right)=0$
and $T$ denotes the space ordering of the matrix multiplication such
that the $a$ term is always to the left of $a-1$ term. 
The factor $\big(\overline{\mathbf{U}}\,\overline{\mathbf{U}}^{\dagger}\big)^{-1}$ appears due to eq.~\eqref{eq:completeness_flavor} in order to take into account possible nonorthogonality of flavor states. 
Furthermore, $\hat{H}^{\left(a\right)}=\textrm{diag}\left(\lambda_{1}^{\left(a\right)},\lambda_{2}^{\left(a\right)},...,\lambda_{3+N}^{\left(a\right)}\right)$
and $X^{\left(a\right)}$ denote respectively the matrix of eigenvalues
and unitary matrix which diagonalizes $\widetilde{H}$ as $\widetilde{H}^{\left(a\right)}=X^{\left(a\right)\dagger}\hat{H}^{\left(a\right)}X^{\left(a\right)}$
in the interval $x_{a-1}<x<x_{a}$. The neutrino oscillation probability
can be calculated by substituting eq. (\ref{eq:S_x_dependent_potential})
into eq. (\ref{eq:prob}). 

We will end this section by proving the following theorem.
\begin{theorem*}
If $\left(\mathbf{U}\mathbf{U}^{\dagger}\right)_{\alpha\alpha}\neq1$
and $\left(\mathbf{U}\mathbf{U}^{\dagger}\right)_{\alpha\beta}=0$
for all $\alpha\neq\beta$, then the neutrino oscillation probability
in an \emph{arbitrary matter potential} is \emph{indistinguishable}
from the \emph{unitarity} scenario. 
\end{theorem*}
The proof is as follows
\begin{eqnarray}
\sum_{i}\overline{\mathbf{U}}_{\beta i}\overline{\mathbf{U}}_{\alpha i}^{*} & = & \frac{1}{\sqrt{\left(\mathbf{U}\mathbf{U}^{\dagger}\right)_{\alpha\alpha}\left(\mathbf{U}\mathbf{U}^{\dagger}\right)_{\beta\beta}}}\sum_{i}\mathbf{U}_{\beta i}\mathbf{U}_{\alpha i}^{*}=\frac{\left(\mathbf{U}\mathbf{U}^{\dagger}\right)_{\alpha\alpha}\delta_{\alpha\beta}}{\sqrt{\left(\mathbf{U}\mathbf{U}^{\dagger}\right)_{\alpha\alpha}\left(\mathbf{U}\mathbf{U}^{\dagger}\right)_{\beta\beta}}}=\delta_{\alpha\beta}.
\end{eqnarray}
From the above, it follows that $\left(\overline{\mathbf{U}}\,\overline{\mathbf{U}}^{\dagger}\right)^{-1}=\mathbf{I}\implies\overline{\mathbf{U}}^{\dagger,-1}\overline{\mathbf{U}}^{-1}=\mathbf{I}\implies\overline{\mathbf{U}}^{\dagger}\overline{\mathbf{U}}=\mathbf{I}$
and hence $\overline{\mathbf{U}}$ is unitary. With unitary $\overline{\mathbf{U}}$,
the Hamiltonians in the vacuum mass basis in eq. (\ref{eq:H_mass_basis})
reduces and coincides with the unitarity one and hence the solution
in eq. (\ref{eq:S_a}) will coincide with the unitarity scenario as
well. This result holds for an arbitrary matter potential $V\left(x\right)$
since one can always construct the full solution as in (\ref{eq:S_x_dependent_potential}).
Finally, we will also recover eq.~\eqref{eq:orthogonal_op} as will be discussed in Section~\ref{sec:osc_prob_nonorthogonal}. 
Let us denote this scenario as the \emph{hidden nonunitarity} scenario.

\subsection{Identities}

The combination that appears in the oscillation amplitude in the flavor
basis (\ref{eq:S_constant_V}) is
\begin{eqnarray}
\mathbf{\widetilde{U}}_{\beta k}\mathbf{\widetilde{U}}_{\alpha k}^{*} & \equiv & \sum_{i,j}\overline{\mathbf{U}}_{\beta i}\overline{\mathbf{U}}_{\alpha j}^{*}X_{ik}X_{jk}^{*}.
\end{eqnarray}
Substituting eq. (\ref{eq:main_result}) into the equation above,
we obtain
\begin{eqnarray}
\mathbf{\widetilde{U}}_{\beta k}\mathbf{\widetilde{U}}_{\alpha k}^{*} & = & \frac{{\displaystyle \sum_{p=0}^{2+N}}\left(-1\right)^{p}\left(\overline{H}^{p}\right)_{\beta\alpha}c_{2+N-p,k}}{Z_{k}},
\end{eqnarray}
where we have defined
\begin{eqnarray}
\overline{H} & \equiv & \overline{\mathbf{U}}\widetilde{H}\overline{\mathbf{U}}^{\dagger}=H\overline{\mathbf{U}}\,\overline{\mathbf{U}}^{\dagger}=\overline{\mathbf{U}}\Delta\overline{\mathbf{U}}^{\dagger}+V,\label{eq:H_intermediate}
\end{eqnarray}
and we have used eq. (\ref{eq:H_mass_basis}) and eq. (\ref{eq:H_flavor_basis})
to arrive at the last equality. It is important to note that $\overline{H}$
is not equal to the Hamiltonian in the flavor basis $H$ but they
only coincide with each other if $\overline{\mathbf{U}}$ is unitary. Furthermore,
$\overline{H}$, not being related to $\widetilde{H}$ and $H$ by
similarity transformation, does not have to have the same eigenvalues as $\widetilde{H}$
and $H$. 

Under trace transformation $H\to H+c\,\mathbf{I}$ with $c$ any real constant or phase transformation $H\to\Phi H\Phi^{\dagger}$ with
$\Phi=\textrm{diag}\left(e^{i\phi_{1}},e^{i\phi_{2}},...\right)$, the probabilities (measureables) (\ref{eq:prob}) remain invariant.
By observing that the following trace and phase transformation invariant
combinations~\cite{Harrison:2002ee}
\begin{eqnarray}
I_{\alpha\beta} & = & \textrm{Im}\left[\left(H^{2}\right)_{\alpha\beta}H_{\alpha\beta}^{*}\right],\label{eq:I_combination}\\
R_{\alpha\beta} & = & \left|H_{\alpha\beta}\right|^{2},\quad\alpha\neq\beta,\label{eq:R_combination}
\end{eqnarray}
should be independent of matter potential if the matter potential
is \emph{diagonal} in the flavor basis, several matter invariant identities can be derived.  
With unitary $\overline{\mathbf{U}}$, the first one results in the Naumov-Harrison-Scott (NHS) identity \cite{Naumov:1991ju,Harrison:2002ee} or their generalized versions \cite{Abdullahi:2022fkh} while the second one results in further matter-invariant identities \cite{Harrison:2002ee,Abdullahi:2022fkh}. In the nonunitarity
scenario, this is no longer true since the matter potential in the
flavor basis (\ref{eq:H_flavor_basis}) is no longer diagonal but
given by $V\left(\overline{\mathbf{U}}\,\overline{\mathbf{U}}^{\dagger}\right)^{-1}$.
Incidentally, this also shows that once we have nondiagonal NonStandard neutrino Interaction (NSI), eqs.
(\ref{eq:I_combination}) and (\ref{eq:R_combination}) are no longer
invariant under matter potential~\cite{Fong:2022oim,Abdullahi:2022fkh}. 
As shown in ref.~\cite{Fong:2022oim}, NSI is still distinct from nonunitarity scenario since the latter further breaks the unitarity relations that we will discuss next.

Let us define the Jarlskog combinations by taking the imaginary part
of the combination above~\cite{Jarlskog:1985ht}
\begin{eqnarray}
J_{\beta\alpha}^{jk} & \equiv & \textrm{Im}\left(\mathbf{\widetilde{U}}_{\beta j}\mathbf{\widetilde{U}}_{\alpha j}^{*}\mathbf{\widetilde{U}}_{\beta k}^{*}\mathbf{\widetilde{U}}_{\alpha k}\right).\label{eq:Jarlskog_combinations}
\end{eqnarray}
If $\widetilde{\mathbf{U}}$ is unitary, one must have
\begin{eqnarray}
\sum_{k}J_{\beta\alpha}^{jk} & = & \sum_{k}\textrm{Im}\left(\mathbf{\widetilde{U}}_{\beta j}\mathbf{\widetilde{U}}_{\alpha j}^{*}\mathbf{\widetilde{U}}_{\beta k}^{*}\mathbf{\widetilde{U}}_{\alpha k}\right)=\textrm{Im}\left(\mathbf{\widetilde{U}}_{\beta j}\mathbf{\widetilde{U}}_{\beta j}^{*}\right)=0.\label{eq:unitary_relations}
\end{eqnarray}
Let us look at the modification due to nonunitarity. Since all the
terms with $p=q$ are real and we are only left with terms of $p\neq q$
\begin{eqnarray}
J_{\beta\alpha}^{jk} & = & {\displaystyle \sum_{p\neq q;0\leq p<q}^{2+N}}\left(-1\right)^{p+q}
\frac{c_{2+N-p,j}c_{2+N-q,k}-c_{2+N-q,j}c_{2+N-p,k}}{Z_j Z_k}
\textrm{Im}\left[\left(\overline{H}^{p}\right)_{\beta\alpha}\left(\overline{H}^{q}\right)_{\beta\alpha}^{*}\right],
\end{eqnarray}
where there is no sum over $j$ and $k$ on the right and we have
used $\textrm{Im}z=-\textrm{Im}z^{*}$. Summing over $k$ and making use of
\begin{eqnarray}
\sum_{k=1}^{3+N}\frac{c_{2+N-p,k}}{Z_{k}} & = & \begin{cases}
0 & p>0\\
1 & p=0
\end{cases},
\end{eqnarray}
we arrive at
\begin{eqnarray}
\sum_{k}J_{\beta\alpha}^{jk} & = & {\displaystyle \sum_{q=1}^{2+N}}\left(-1\right)^{q+1} \frac{c_{2+N-q,j}}{Z_j}
\textrm{Im}\left[\big(\overline{\mathbf{U}}\,\overline{\mathbf{U}}^{\dagger}\big)_{\beta\alpha}\left(\overline{H}^{q}\right)_{\beta\alpha}^{*}\right],\label{eq:general_identity}
\end{eqnarray}
where we have used $\overline{H}^{0}=\overline{\mathbf{U}}\,\overline{\mathbf{U}}^{\dagger}$.
For unitary $\overline{\mathbf{U}}\implies\big(\overline{\mathbf{U}}\,\overline{\mathbf{U}}^{\dagger}\big)_{\beta\alpha}=\delta_{\beta\alpha}$
and since $\overline{H}^{q}$ is Hermitian, the diagonal elements
are real and we recover eq. (\ref{eq:unitary_relations}). This is
consistent with the theorem we have proven and indeed, if $\left(\mathbf{U}\mathbf{U}^{\dagger}\right)_{\alpha\beta}=0$
for all $\alpha\neq\beta$, the right hand side of eq. (\ref{eq:general_identity})
vanishes. In the vacuum, $\overline{H}^{q}=\big(\overline{\mathbf{U}}\Delta\overline{\mathbf{U}}^{\dagger}\big)^{q}$
and one recover
\begin{eqnarray}
\sum_{k}J_{\beta\alpha}^{jk} & = & -\textrm{Im}\left[\big(\overline{\mathbf{U}}\,\overline{\mathbf{U}}^{\dagger}\big)_{\beta\alpha}\overline{\mathbf{U}}_{\alpha j}\overline{\mathbf{U}}_{\beta j}^{*}\right],\label{eq:general_identity_vacuum}
\end{eqnarray}
as can also be derived directly from eq. (\ref{eq:Jarlskog_combinations}).
So by measuring these relations above, we can uncover unitarity violation
in the matter (\ref{eq:general_identity}) or in the vacuum (\ref{eq:general_identity_vacuum}).

\section{Oscillation probability for nonorthogonal flavor states}\label{sec:osc_prob_nonorthogonal}

Probability is not an observable in quantum mechanics and there is no associated Hermitian operator. Typically, to calculate probability of a state $\left|\nu_\alpha\right\rangle$ being found in $\left|\nu_\beta\right\rangle$, one insert the projection operator $\left|\nu_\beta\right\rangle\left\langle\nu_\beta\right|$ in between $\left\langle \nu_{\alpha}|\nu_{\alpha}\right\rangle$ and obtain the probability $P_{\beta\alpha} = |\left\langle \nu_{\beta}|\nu_{\alpha}\right\rangle|^2$ which is the Born rule.

When the complete set of states $\left\{ \left|\nu_\alpha\right\rangle \right\} $
are not orthogonal, from eq.~\eqref{eq:completeness_flavor}, the projection operator becomes
\begin{eqnarray}
	P_{\alpha} & \equiv & \sum_{\beta}  \left|\nu_\alpha\right\rangle \left[\big(\overline{\mathbf{U}}\, \overline{\mathbf{U}}^\dagger\big)^{-1} \right]_{\alpha\beta} \left\langle \nu_\beta \right|,
	\label{eq:general_projection}
\end{eqnarray}
which satisfies $P_{\alpha}^{2}=P_{\alpha}$ and $ \sum_{\alpha}P_{\alpha}=\mathbf{I}$. 
Inserting this projection operator in between $\left\langle \nu_{\alpha}|\nu_{\alpha}\right\rangle$, we obtain
\begin{eqnarray}
	\left\langle \nu_\alpha|P_{\beta}|\nu_\alpha\right\rangle  & = & \left|\left\langle \nu_{\beta}|\nu_{\alpha}\right\rangle\right|^{2}
	+\sum_{\gamma\neq\beta}
	\left\langle \nu_{\alpha}|\nu_{\beta}\right\rangle
	\left[\big(\overline{\mathbf{U}}\, \overline{\mathbf{U}}^\dagger\big)^{-1} \right]_{\beta\gamma} 
	\left\langle \nu_{\gamma}|\nu_{\alpha}\right\rangle.
\end{eqnarray}
Notice that the second term is in general complex and hence the quantity above cannot be interpreted as a probability. Besides being
real and positive, one has to make sure that the probabilities of
finding $\left|\nu_\alpha\right\rangle $ in all possible $\left|\nu_\beta\right\rangle $
sum up to \emph{unity}. According to the theorem in the last section, if $\left(\mathbf{U}\mathbf{U}^{\dagger}\right)_{\alpha\beta}=0$ for all $\alpha\neq\beta$, eq.~\eqref{eq:general_projection} becomes the standard projector and we recover eq.~\eqref{eq:orthogonal_op}.

Nonorthogonal basis states are commonplace in quantum chemistry~\cite{Mulliken:1955,Roby:1974,Leon:1988,Manning:1990,Artacho:1991,Soriano:2014,Artacho:2017}, for example, to express molecular orbitals as linear combinations of atomic orbitals which are in general not orthogonal. In particle physics, nonorthogonal basis states can arise due to new physics as in our consideration of high scale nonunitarity scenario.
In ref. \cite{Leon:1988,Manning:1990}, the theory of projected probabilities
on nonorthogonal states are developed and for two and three states system, the probability operators have closed form. We will write down the results here and defer the details to Appendix~\ref{app:probability_operator}.

Before moving to the realistic three-flavor scenario, we will first show the two-flavor result since they are simpler and illustrative.
Denoting ${\cal N}_{\alpha\beta}\equiv(\overline{U}\,\overline{U}^{\dagger})_{\alpha\beta}$ and let us choose the two flavors as $e,\mu$ (it can also be $e,\tau$ or $\mu, \tau$), $(\hat{P}_{\alpha})_{\xi\lambda}$ in eq.~\eqref{eq:prob} is given by
\begin{eqnarray}
	(\hat{P}_{\alpha})_{\xi\lambda} & = & \begin{cases}
		\displaystyle{1 +
		\frac{\left|{\cal N}_{e\mu}\right|^{4}}{1-\left|{\cal N}_{e\mu}\right|^{4}}}, & \xi=\lambda=\alpha\\
		{\displaystyle \frac{\left|{\cal N}_{e\mu}\right|^{2}}{1-\left|{\cal N}_{e\mu}\right|^{4}}}, & \xi=\lambda\neq\alpha\\
		\displaystyle{-\frac{1}{2}\frac{{\cal N}_{\xi\lambda}}{1-\left|{\cal N}_{e\mu}\right|^{2}}}, & \xi\neq\lambda\;\textrm{and}\;\xi=\alpha\;\textrm{or}\;\lambda=\alpha
	\end{cases}.
\end{eqnarray}
In the case where off-diagonal elements of ${\cal N}$ are zero, we recover eq.~\eqref{eq:orthogonal_op}. It is clear that the deviation from the standard unitarity scenario depends on the off-diagonal elements. From eq.~\eqref{eq:prob}, let us write down the oscillation probabilities explicitly
\begin{eqnarray}
	P_{ee} & = & \left|S_{ee}\right|^{2}+\frac{\left|{\cal N}_{e\mu}\right|^{4}\left|S_{ee}\right|^{2}+\left|{\cal N}_{e\mu}\right|^{2}\left|S_{e\mu}\right|^{2}}{1-\left|{\cal N}_{e\mu}\right|^{4}}-\frac{\textrm{Re}\left(S_{ee}{\cal N}_{e\mu}S_{\mu e}\right)}{1-\left|{\cal N}_{e\mu}\right|^{2}},\\
	P_{\mu e} & = & \left|S_{e\mu}\right|^{2}+\frac{\left|{\cal N}_{e\mu}\right|^{4}\left|S_{e\mu}\right|^{2}+\left|{\cal N}_{e\mu}\right|^{2}\left|S_{ee}\right|^{2}}{1-\left|{\cal N}_{e\mu}\right|^{4}}-\frac{\textrm{Re}\left(S_{ee}{\cal N}_{e\mu}S_{\mu e}\right)}{1-\left|{\cal N}_{e\mu}\right|^{2}},\\
	P_{e\mu} & = & \left|S_{e\mu}\right|^{2}+\frac{\left|{\cal N}_{e\mu}\right|^{4}\left|S_{e\mu}\right|^{2}+\left|{\cal N}_{e\mu}\right|^{2}\left|S_{\mu\mu}\right|^{2}}{1-\left|{\cal N}_{e\mu}\right|^{4}}-\frac{\textrm{Re}\left(S_{\mu\mu}{\cal N}_{\mu e}S_{e\mu}\right)}{1-\left|{\cal N}_{e\mu}\right|^{2}},\\
	P_{\mu\mu} & = & \left|S_{\mu\mu}\right|^{2}+\frac{\left|{\cal N}_{e\mu}\right|^{4}\left|S_{\mu\mu}\right|^{2}+\left|{\cal N}_{e\mu}\right|^{2}\left|S_{e\mu}\right|^{2}}{1-\left|{\cal N}_{e\mu}\right|^{4}}-\frac{\textrm{Re}\left(S_{\mu\mu}{\cal N}_{\mu e}S_{e\mu}\right)}{1-\left|{\cal N}_{e\mu}\right|^{2}}.
\end{eqnarray}
The first terms have the standard form, while the additional terms are required to ensure unitarity. Notice that the expressions above hold for any matter potential (including vacuum) since the dynamics is contained in the amplitude $S$ obtained by solving the Schrödinger equation in Section \ref{subsec:evolution_flavor_state}. 
We can verify explicitly that unitarity is preserved
\begin{eqnarray}
	P_{ee}+P_{\mu e} & = & \frac{\left|S_{ee}\right|^{2}+\left|S_{e\mu}\right|^{2}-2\textrm{Re}\left(S_{ee}{\cal N}_{e\mu}S_{\mu e}\right)}{1-\left|{\cal N}_{e\mu}\right|^{2}}=\sum_{\alpha,\beta}\left\langle \nu_{e}|\nu_{\alpha}\right\rangle \left({\cal N}^{-1}\right)_{\alpha\beta}\left\langle \nu_{\beta}|\nu_{e}\right\rangle =1,\\
	P_{e\mu}+P_{\mu\mu} & = & \frac{\left|S_{\mu\mu}\right|^{2}+\left|S_{e\mu}\right|^{2}-2\textrm{Re}\left(S_{\mu\mu}{\cal N}_{\mu e}S_{e\mu}\right)}{1-\left|{\cal N}_{e\mu}\right|^{2}}=\sum_{\alpha,\beta}\left\langle \nu_{\mu}|\nu_{\alpha}\right\rangle \left({\cal N}^{-1}\right)_{\alpha\beta}\left\langle \nu_{\beta}|\nu_{\mu}\right\rangle =1,
\end{eqnarray}
where we have used eq.~\eqref{eq:completeness_flavor} in the third equalities.

For three flavors scenario $\alpha,\beta,...=\{e,\mu,\tau\}$, ${\cal N}$ is a $3\times 3$ matrix with diagonal elements all equal to one. We will next define ${\cal N}_\alpha$ as a $2\times 2$ submatrix formed from the matrix ${\cal N}$ excluding the row and column involving $\nu_\alpha$ state.
The result is
\begin{eqnarray}
	(\hat{P}_{\alpha})_{\xi\lambda} & = & \frac{1}{3}\left[(E_{\alpha})_{\xi\lambda}
	+\sum_{\beta\neq\alpha}(F_{\alpha\beta})_{\xi\lambda}\right],
\end{eqnarray}
with
\begin{eqnarray}
	\left(E_{\alpha}\right)_{\xi\lambda} & = & \begin{cases}
		{\displaystyle 1 + \frac{X_{\alpha}^{2}}{1-X_{\alpha}^{2}}}, & \xi=\lambda=\alpha\\
		{\displaystyle \frac{\left|{\cal N}_{\alpha\xi}-{\cal N}_{\alpha\gamma\xi}\right|^{2}}{\left(\det {\cal N}_{\alpha}\right)^{2}\left(1-X_{\alpha}^{2}\right)},\;\gamma\neq\left\{ \alpha,\xi\right\} }, & \xi=\lambda\neq\alpha\\
		{\displaystyle -\frac{1}{2}\frac{{\cal N}_{\xi\lambda}-{\cal N}_{\xi\gamma\lambda}}{\det {\cal N}}},\;\gamma\neq\left\{ \alpha,\xi\right\} , & \xi\neq\lambda\;\textrm{and}\;(\xi=\alpha\;\textrm{or}\;\lambda=\alpha)\\
		{\displaystyle \frac{\left({\cal N}_{\alpha\lambda}-{\cal N}_{\alpha\xi\lambda}\right)\left({\cal N}_{\xi\alpha}-{\cal N}_{\xi\lambda\alpha}\right)}{\left(\det {\cal N}_{\alpha}\right)^{2}\left(1-X_{\alpha}^{2}\right)}}, & \xi\neq\lambda\;\textrm{and}\;\left\{ \xi,\lambda\right\} \neq\alpha
	\end{cases},
\end{eqnarray}
and for $\beta\neq\alpha$ and $\gamma\neq\left\{ \alpha,\beta\right\} $
\begin{eqnarray}
	\left(F_{\alpha\beta}\right)_{\xi\lambda} & = & \begin{cases}
		{\displaystyle 1 + \frac{\left|{\cal N}_{\alpha\gamma}\right|^{4}}{1-\left|{\cal N}_{\alpha\gamma}\right|^{4}}
			+\frac{\langle\left(p_{\alpha}\right)_{\left\{ \alpha,\gamma\right\} }\rangle_{\beta\beta} \left|{\cal N}_{\alpha\beta}-{\cal N}_{\alpha\gamma\beta}\right|^{2}}{\left(\det {\cal N}_{\beta}\right)^{2}\left(1-X_{\beta}^{2}\right)}}, & \xi=\lambda=\alpha\\
		{\displaystyle \begin{split}-\frac{1}{2}\frac{1}{1-\left|{\cal N}_{\alpha\gamma}\right|^{4}}\left({\cal N}_{\xi\lambda}-\frac{1+\left|{\cal N}_{\alpha\gamma}\right|^{2}}{2}{\cal N}_{\xi\gamma\lambda}\right)\\
				-\frac{1}{2}\frac{\langle\left(p_{\alpha}\right)_{\left\{ \alpha,\gamma\right\} }\rangle_{\beta\beta} }{1-X_{\beta}^{2}}
				\frac{{\cal N}_{\xi\lambda}-{\cal N}_{\xi\gamma\lambda}}{\det {\cal N}_{\beta}}\left(1+\left\langle p_{\alpha\gamma}\right\rangle_{\beta\beta} \right)
			\end{split}
		}, & \xi\lambda=\alpha\beta\;\textrm{or}\;\xi\lambda=\beta\alpha\\
		{\displaystyle 
			-\frac{1}{2}\frac{{\cal N}_{\xi\lambda}}{\det {\cal N}_{\beta}}
			+\frac{\langle\left(p_{\alpha}\right)_{\left\{ \alpha,\gamma\right\} }\rangle_{\beta\beta} \left({\cal N}_{\xi\beta}-{\cal N}_{\xi\lambda\beta}\right)\left({\cal N}_{\beta\lambda}-{\cal N}_{\beta\xi\lambda}\right)}{\left(\det {\cal N}_{\beta}\right)^{2}\left(1-X_{\beta}^{2}\right)}}, & \xi\lambda=\alpha\gamma\;\textrm{or}\;\xi\lambda=\gamma\alpha\\
		{\displaystyle \frac{1}{2}\left(1+\frac{1+\left\langle p_{\alpha\gamma}\right\rangle_{\beta\beta}^{2}}{1-X_{\beta}^{2}}\right)
			\langle\left(p_{\alpha}\right)_{\left\{ \alpha,\gamma\right\} }\rangle_{\beta\beta} }, & \xi=\lambda=\beta\\
		{\displaystyle \begin{split}-\frac{1}{2}\frac{1}{1-\left|{\cal N}_{\alpha\gamma}\right|^{4}}\left(\left|{\cal N}_{\alpha\gamma}\right|^{2}{\cal N}_{\xi\lambda}-\frac{1+\left|{\cal N}_{\alpha\gamma}\right|^{2}}{2}{\cal N}_{\xi\alpha\lambda}\right)\\
				-\frac{1}{2}\frac{\langle\left(p_{\alpha}\right)_{\left\{ \alpha,\gamma\right\} }\rangle_{\beta\beta} }{1-X_{\beta}^{2}}
				\frac{{\cal N}_{\xi\lambda}-{\cal N}_{\xi\alpha\lambda}}{\det {\cal N}_{\beta}}\left(1+\left\langle p_{\alpha\gamma}\right\rangle_{\beta\beta} \right)
			\end{split}
			,} & \xi\lambda=\beta\gamma\;\textrm{or}\;\xi\lambda=\gamma\beta\\
		{\displaystyle \frac{\left|{\cal N}_{\alpha\gamma}\right|^{2}}{1-\left|{\cal N}_{\alpha\gamma}\right|^{4}}
			+\frac{\langle\left(p_{\alpha}\right)_{\left\{ \alpha,\gamma\right\} }\rangle_{\beta\beta} 
				\left|{\cal N}_{\gamma\beta}-{\cal N}_{\gamma\alpha\beta}\right|^{2}}{\left(\det {\cal N}_{\beta}\right)^{2}\left(1-X_{\beta}^{2}\right)}}, & \xi=\lambda=\gamma
	\end{cases},
\end{eqnarray}
where we have defined
\begin{eqnarray}
	{\cal N_{\alpha\beta\gamma}} & \equiv &
	{\cal N}_{\alpha\beta}{\cal N}_{\beta\gamma}, \\
	\left\langle p_{\alpha\gamma}\right\rangle_{\beta\beta}  & \equiv & \frac{\left|{\cal N}_{\alpha\beta}\right|^{2}+\left|{\cal N}_{\beta\gamma}\right|^{2}
		-2\textrm{Re}\left({\cal N}_{\beta\alpha}{\cal N}_{\alpha\gamma}{\cal N}_{\gamma\beta}\right)}{1-\left|{\cal N}_{\alpha\gamma}\right|^{2}},\\
	\langle\left(p_{\alpha}\right)_{\left\{ \alpha,\gamma\right\} }\rangle_{\beta\beta}  & \equiv & \frac{\left|{\cal N}_{\alpha\beta}\right|^{2}
		+\left|{\cal N}_{\alpha\gamma}{\cal N}_{\beta\gamma}\right|^{2}
		-(1+\left|{\cal N}_{\alpha\gamma}\right|^{2})
		\textrm{Re}\left({\cal N}_{\beta\alpha}{\cal N}_{\alpha\gamma}{\cal N}_{\gamma\beta}\right)}{1-\left|{\cal N}_{\alpha\gamma}\right|^{4}}.
\end{eqnarray}
The subscript $\{\alpha,\gamma\}$ in the second expression are not indices but refers to the set of basis states of the corresponding operator. For example, with $\{\mu,\tau\}$, we can have $\langle\left(p_{\mu}\right)_{\left\{ \mu,\tau\right\} }\rangle_{\beta\beta}$ or $\langle\left(p_{\tau}\right)_{\left\{ \mu,\tau\right\} }\rangle_{\beta\beta}$.
In the case where off-diagonal elements of ${\cal N}$ are zero, we again recover eq.~\eqref{eq:orthogonal_op}.

\section{High versus low scale nonunitarity}\label{sec:high_vs_low}

\subsection{In vacuum}

In the absence of matter $X_{ij}=\delta_{ij}$, eq.~\eqref{eq:prob}
becomes
\begin{eqnarray}
	P_{\beta\alpha}\left(x\right) & = &
    \sum_{i,j}\sum_{\xi,\lambda}
    \overline{\mathbf{U}}_{\alpha i}\overline{\mathbf{U}}_{\xi i}^{*}e^{\frac{im_{i}^{2}x}{2E}}
    (\hat P_{\beta})_{\xi\lambda}
	 \overline{\mathbf{U}}_{\lambda j}\overline{\mathbf{U}}_{\alpha j}^{*}e^{-\frac{im_{j}^{2}x}{2E}}.
	 \label{eq:P_vacuum}
\end{eqnarray}
For the high scale nonunitarity scenario, $\left(\overline{\mathbf{U}}\,\overline{\mathbf{U}}^{\dagger}\right)_{\alpha\beta}=\left(\overline{U}\,\overline{U}^{\dagger}\right)_{\alpha\beta}$ where $U$ spans over three flavors and one can write
\begin{eqnarray}
	P_{\beta\alpha}\left(x\right) & = &
	\sum_{i,j}\sum_{\xi,\lambda}
	\overline{U}_{\alpha i}\overline{U}_{\xi i}^{*}e^{\frac{im_{i}^{2}x}{2E}}
	(\hat P_{\beta})_{\xi\lambda}
	\overline{U}_{\lambda j}\overline{U}_{\alpha j}^{*}e^{-\frac{im_{j}^{2}x}{2E}}.
	\label{eq:P_high}
\end{eqnarray} 
According to the theorem proved
in the previous section, in the hidden nonunitarity scenario $\left(UU^{\dagger}\right)_{\alpha\beta}=0$
for all $\alpha\neq\beta$, one cannot distinguish it from unitarity
scenario since $\overline{U}$ will be unitary. This implies that
nonunitarity effect is proportional to $\left(UU^{\dagger}\right)_{\alpha\beta}$
for $\alpha\neq\beta$ as encapsulated in eq. (\ref{eq:general_identity_vacuum}),
making it more challenging to distinguish it from the unitarity scenario if the modulus of $\left(UU^{\dagger}\right)_{\alpha\beta}$ is small (this conclusion also holds in an arbitrary matter potential as we
will discuss in the next subsection).

Let us contrast the high scale nonunitarity scenario to the low
scale nonunitarity scenario.
\begin{enumerate}
\item[(i)]  For the low scale nonunitarity scenario, flavor states remain orthogonal, $\mathbf{U}$ is unitary
with $\left(\mathbf{U}\mathbf{U}^{\dagger}\right)_{\alpha\alpha}=1$
and we have
\begin{eqnarray}
P_{\beta\alpha}^{\textrm{low}}(x) & = & \left|\sum_{i=1}^{3+N}\mathbf{U}_{\beta i}\mathbf{U}_{\alpha i}^{*}e^{-\frac{im_{i}^{2}x}{2E}}\right|^{2}.\label{eq:P_low}
\end{eqnarray}
The most direct way to discover low scale nonunitarity scenario is to have an experiment with $m_{i>3}^{2}\sim E/x$
such that oscillations involving new fermions can be measured.\footnote{For example, a recent short baseline reactor experiment STEREO \cite{STEREO:2022nzk} rules out the existence of a sterile neutrino with mass in the eV range and mixing element of the order of 0.4 and larger.} The main challenge is, a priori, we do not know $m_{i>3}$ or even if
$\nu_{i>3}$ exist but one can design an experiment with identical
detectors at various baselines to cover as large range of $E/x$ as possible. 
In the scenario, unitarity relations (\ref{eq:unitary_relations})
are satisfied in contrast to high scale nonunitarity scenario which satisfies
(\ref{eq:general_identity_vacuum}).
While $P_{\beta\alpha}^{\textrm{low}}(0) = \delta_{\beta\alpha}$, there is a \emph{zero distance} effect for the high scale nonunitarity scenario
\begin{eqnarray}
	P_{\beta\alpha}^{\textrm{high}}(0) & = &
	\sum_{\xi,\lambda}
	(\overline{U}\,\overline{U}^{\dagger})_{\alpha\xi}
	(\hat P_{\beta})_{\xi\lambda}
	(\overline{U}\,\overline{U}^{\dagger})_{\lambda\alpha}.
	\label{eq:P_high_zero}
\end{eqnarray}

\item[(ii)]  If $m_{i>3}\gg \sqrt{\Delta m_\textrm{atm}^2}$ such that the oscillations involving them are averaged out, we obtain~\cite{Fong:2016yyh,Fong:2017gke}
\begin{eqnarray}
P_{\beta\alpha}^{\textrm{low,ave}}(x) & = & {\cal C}_{\alpha\beta}+\left|\sum_{i=1}^{3}\hat{U}_{\beta i}\hat{U}_{\alpha i}^{*}e^{-\frac{im_{i}^{2}x}{2E}}\right|^{2},\label{eq:P_low_ave}
\end{eqnarray}
where we use $\hat{U}$ to denote $3\times 3$ submatrix from $\mathbf{U}$ and
${\cal C}_{\alpha\beta}$ is an additional constant term, also known
as the \emph{probability leaking term}
\begin{eqnarray}
{\cal C}_{\alpha\beta} & = & \sum_{i=4}^{3+N}\left|\mathbf{U}_{\alpha i}\right|^{2}\left|\mathbf{U}_{\beta i}\right|^{2}.
\end{eqnarray}
It is bounded from above and below~\cite{Fong:2016yyh}
\begin{eqnarray}
\frac{1}{N}d_{\alpha\beta}\leq & {\cal C}_{\alpha\beta} & \leq d_{\alpha\beta},
\end{eqnarray}
where $d_{\alpha\beta}\equiv \left(1-\sum_{i=1}^{3}\left|\hat{U}_{\alpha i}\right|^{2}\right)\left(1-\sum_{j=1}^{3}\left|\hat{U}_{\beta j}\right|^{2}\right)$.
In principle, a measurement of ${\cal C}_{\alpha\beta}$ will allow
us to obtain information about the number $N$ of additional fermions.
Notice that for $N=1$, ${\cal C}_{\alpha\beta}$ is completely fixed
by $\hat{U}$. In practice, it is a challenging task since this term
is expected to be small, being fourth order in unitarity violation
parameter
\begin{eqnarray}
\epsilon_{\alpha\beta} & \equiv & \sqrt{\left|\delta_{\alpha\beta}-\sum_{i=1}^{3}\hat{U}_{\beta i}\hat{U}_{\alpha i}^{*}\right|}.
\end{eqnarray}
Besides the normalization factor $\sqrt{\left(UU^{\dagger}\right)_{\alpha\alpha}}$ in eq.~\eqref{eq:P_high}, we have 
nontrivial structure in $\hat P_\beta$ due to nonorthogonality of flavor states. In principle, this allows us to distinguish between the two scenarios. 
Furthermore, one can also measure the normalization factor in electroweak precision measurements~\cite{Antusch:2014woa,Fernandez-Martinez:2016lgt}. As we will see next, in the presence of matter, one will have new nontrivial effect.

\end{enumerate}

\begin{figure}
\begin{centering}
\includegraphics[scale=0.65]{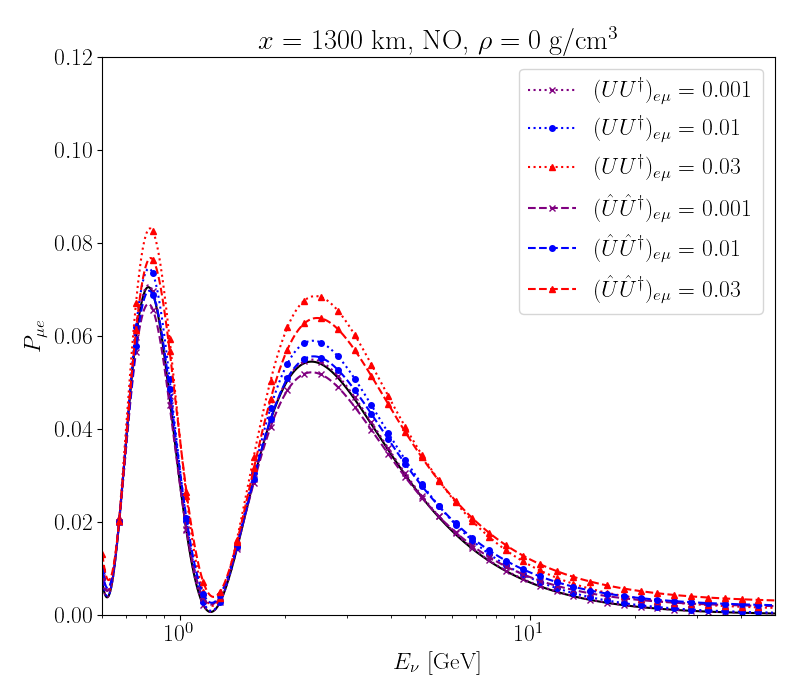}
\par\end{centering}
\caption{Comparison of the probability of $\nu_\mu \to \nu_e$ at $x=1300$ km in the vacuum as a function of neutrino energy $E_\nu$ between the high scale nonunitarity scenario (dotted lines) and low scale nonunitarity scenario (dashed lines) with $(UU^{\dagger})_{ee}=(UU^{\dagger})_{\mu\mu}=(\hat{U}\hat{U}^{\dagger})_{ee}=(\hat{U}\hat{U}^{\dagger})_{\mu\mu}=0.96$. The solid black line is the standard three-flavor unitarity scenario. \label{fig:nonunitarity_vacuum}}
\end{figure}

\begin{figure}
\begin{centering}
\includegraphics[scale=0.65]{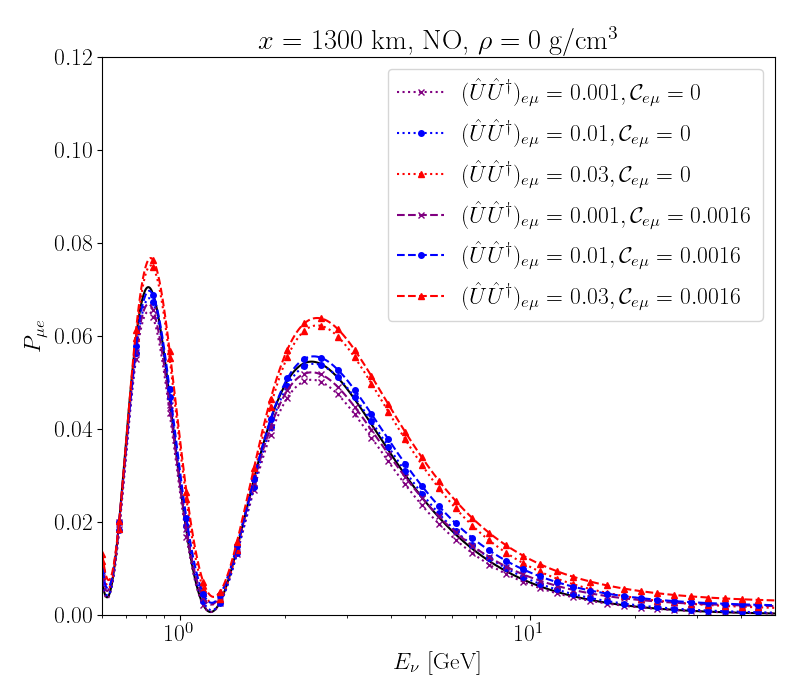}
\par\end{centering}
\caption{The probability of $\nu_\mu \to \nu_e$ at $x=1300$ km in the vacuum as a function of neutrino energy $E_\nu$ for low scale nonunitarity scenario with $(\hat{U}\hat{U}^{\dagger})_{ee}=(\hat{U}\hat{U}^{\dagger})_{\mu\mu}=0.96$ by setting the leaking term to be the maximum ${\cal C}_{e\mu}=0.0016$ (dashed lines) and the minimum ${\cal C}_{e\mu}=0$ (dotted lines). The solid black line is the standard three-flavor unitarity scenario.\label{fig:nonunitarity_vacuum_leaking}}
\end{figure}

In Figure \ref{fig:nonunitarity_vacuum}, we plot the oscillation
probability for $\nu_{\mu}\to\nu_{e}$ as a function of neutrino energy
$E_{\nu}$ fixing the baseline $x=1300$ km, for the standard three-flavor
unitarity scenario (solid black line), high scale (dotted lines) and
low scale (dashed lines) nonunitarity scenarios. 
Here and in the following, the standard parameters
are always set to the global best fit values for NO from~\cite{Esteban:2020cvm,NuFIT}. For the high scale nonunitarity scenario, we set $\left(UU^{\dagger}\right)_{ee}=\left(UU^{\dagger}\right)_{\mu\mu}=0.96$, $\left(UU^{\dagger}\right)_{\tau\tau} = 1$, $\left(UU^{\dagger}\right)_{e\tau} = \left(UU^{\dagger}\right)_{\mu\tau} = 0$ and $\left(UU^{\dagger}\right)_{e\mu}=\left\{ 10^{-3},10^{-2},0.03\right\} $.
To compare with the low scale nonunitarity scenario where oscillations
involving $\nu_{i>3}$ are averaged out, we also set $(\hat{U}\hat{U}^{\dagger})_{ee}=(\hat{U}\hat{U}^{\dagger})_{\mu\mu}=0.96$, $(\hat U \hat U^{\dagger})_{\tau\tau} = 1$, $(\hat U \hat U^{\dagger})_{e\tau} = (\hat U \hat U^{\dagger})_{\mu\tau} = 0$
and $(\hat{U}\hat{U}^{\dagger})_{e\mu}=\left\{ 10^{-3},10^{-2},0.03\right\} $.
With this choice, the leaking term is bounded as $0.0016/N<{\cal C}_{e\mu}\leq0.0016$
and we have set ${\cal C}_{\alpha\beta}=0.0016$ for illustration. To illustrate the effect of ${\cal C}_{e\mu}$, we plot in Figure
\ref{fig:nonunitarity_vacuum_leaking} setting ${\cal C}_{e\mu}=0$
(dotted lines) in comparison to the case with ${\cal C}_{e\mu}=0.0016$
(dashed lines).
As we can see explicitly in Figure \ref{fig:nonunitarity_vacuum}, as $\left(UU^{\dagger}\right)_{e\mu}$ decreases,
the high scale nonunitarity scenario approaches the unitarity scenario
while for the low scale nonunitarity scenario, this does not happens.

In the neutrino experiments, the number of observed of neutrinos at
the detector can be written as\footnote{This is a theorist's expression that we have not included experimental
effects like detection efficiency and energy reconstruction.}
\begin{eqnarray}
N_{\beta\alpha} & = & \sigma_{\beta}P_{\beta\alpha}\left(x\right)\phi_{\alpha},
\end{eqnarray}
where $\phi_{\alpha}$ is the $\nu_{\alpha}$ neutrino flux at production
and $\sigma_{\beta}$ is the detection cross section of $\nu_{\beta}$
and the energy dependence of all the terms are left implicit. In order
to determine $P_{\beta\alpha}\left(x\right)$ from $N_{\beta\alpha}$,
it is crucial to have precise determination of $\sigma_{\beta}\times\phi_{\alpha}$.
To mitigate the uncertainty in flux determination, one can take the
ratio of measurement in a far detector placed at $x_{1}$ and a near
detector placed at $x_{0}$ 
\begin{eqnarray}
\frac{N_{\beta\alpha}\left(x_{1}\right)}{N_{\alpha\alpha}\left(x_{0}\right)} & = & \frac{\sigma_{\beta}P_{\beta\alpha}\left(x_{1}\right)\phi_{\alpha}}{\sigma_{\alpha}P_{\alpha\alpha}\left(x_{0}\right)\phi_{\alpha}}=\frac{\sigma_{\beta}P_{\beta\alpha}\left(x_{1}\right)}{\sigma_{\alpha}P_{\alpha\alpha}\left(x_{0}\right)}.
\end{eqnarray}
For the high and low scale nonunitarity scenarios with $x_{0}\ll E/m_{i}^{2}$
for all $i$, eqs. (\ref{eq:P_high}) and (\ref{eq:P_low}) are
\begin{eqnarray}
P_{\alpha\alpha}^{\textrm{high}}\left(x_{0}\right) & \simeq &
(\overline{U}\,\overline{U}^\dagger
\hat P_{\alpha}
\overline{U}\,\overline{U}^\dagger)_{\alpha\alpha}+\mathcal{O}\left(x_{0}^{2}m_{i}^{4}/E^{2}\right), \\
 P_{\alpha\alpha}^{\textrm{low}}\left(x_{0}\right) &\simeq & 1-\mathcal{O}\left(x_{0}^{2}m_{i}^{4}/E^{2}\right),
\end{eqnarray}
which give
\begin{eqnarray}
\left.\frac{N_{\beta\alpha}\left(x_{1}\right)}{N_{\alpha\alpha}\left(x_{0}\right)}\right|_\textrm{high} & \simeq & \frac{\sigma_{\beta}}{\sigma_{\alpha}}
\frac{P_{\beta\alpha}^{\textrm{high}}\left(x_{1}\right)}{(\overline{U}\,\overline{U}^\dagger
	\hat P_{\alpha}
	\overline{U}\,\overline{U}^\dagger)_{\alpha\alpha}}, \label{eq:ratio_high}\\
\left.\frac{N_{\beta\alpha}\left(x_{1}\right)}{N_{\alpha\alpha}\left(x_{0}\right)}\right|_\textrm{low} & \simeq & \frac{\sigma_{\beta}}{\sigma_{\alpha}}P_{\beta\alpha}^{\textrm{low}}\left(x_{1}\right).
\label{eq:ratio_low}
\end{eqnarray}
With dedicated measurements of $\sigma_{\beta}$ and $\sigma_{\alpha}$,
the two scenarios can be distinguished from each other.\footnote{In the high scale nonunitarity scenario, the additional factor which appears
in the cross section in comparison with the SM expectation $\sigma_{\alpha}=\sigma_{\alpha}^{\textrm{SM}}\left(UU^{\dagger}\right)_{\alpha\alpha}$ should already be included in dedicated measurement.} In the low scale nonunitarity scenario with $x_{0}\gg E/m_{i}^{2}$ for
$i>3$ such that the fast oscillations can be averaged out, eq. (\ref{eq:P_low_ave})
gives
\begin{eqnarray}
P_{\alpha\alpha}^{\textrm{low,ave}}\left(x_{0}\right) & \simeq & {\cal C}_{\alpha\alpha}+\left|\sum_{i=1}^{3}\hat{U}_{\alpha i}\hat{U}_{\alpha i}^{*}\right|^{2}-\mathcal{O}\left(x_{0}^{2}m_{i}^{4}/E^{2}\right).
\end{eqnarray}
In this case, one has
\begin{eqnarray}
\left.\frac{N_{\beta\alpha}\left(x_{1}\right)}{N_{\alpha\alpha}\left(x_{0}\right)}\right|_\textrm{low,ave} & \simeq & \frac{\sigma_{\beta}}{\sigma_{\alpha}}\frac{P_{\beta\alpha}^{\textrm{low,ave}}\left(x_{1}\right)}{{\cal C}_{\alpha\alpha}+\left|\sum_{i=1}^{3}\hat{U}_{\alpha i}\hat{U}_{\alpha i}^{*}\right|^{2}}.\label{eq:ratio_low_ave}
\end{eqnarray}
which can be differentiated from eqs.~\eqref{eq:ratio_high} and ~\eqref{eq:ratio_low}. This type of arrangement has been planned in the upcoming neutrino experiments DUNE~\cite{DUNE} and T2HK~\cite{T2HK}.

Besides through neutrino oscillation experiments, the synergy with constraints from the electroweak precision measurements is needed to discover high scale nonunitarity~\cite{Antusch:2014woa,Fernandez-Martinez:2016lgt} or low scale nonunitarity~\cite{deGouvea:2015euy}. If $\left(UU^{\dagger}\right)_{ee}=\left(UU^{\dagger}\right)_{\mu\mu}=\left(UU^{\dagger}\right)_{\tau\tau}\neq1$, one can measure this deviation by comparing the leptonic weak processes with the hadronic weak processes. For instance, while absolute lifetimes of $\mu^{\pm}$, $\pi^{\pm}$, $K^{\pm}$ and $K^{0}$ will be affected, the leptonic
branching ratios will be the same. For $K^{\pm}$ and $K^{0}$, the branching ratios to hadronic and leptonic channels will be modified.
If $\left(UU^{\dagger}\right)_{ee}\neq\left(UU^{\dagger}\right)_{\mu\mu}\neq\left(UU^{\dagger}\right)_{\tau\tau}\neq1$,
lepton universality is broken and one can measure this by studying different leptonic weak processes. The reader can refer to refs.~\cite{Antusch:2014woa,Fernandez-Martinez:2016lgt,deGouvea:2015euy} for more details.

\subsection{In matter}

Now let us consider the scenario with matter effect. For high scale
nonunitarity scenario, eq. (\ref{eq:H_mass_basis}) is just
\begin{eqnarray}
\widetilde{H}^{\textrm{high}} & = & \Delta+\overline{U}^{-1}V\overline{U}^{\dagger,-1},
\end{eqnarray}
where $\overline{U}$ spans over 3 flavors. According to theorem proved
earlier, in the hidden nonunitarity scenario $\left(UU^{\dagger}\right)_{\alpha\beta}=0$
for all $\alpha\neq\beta$, $\overline{U}$ is unitary and hence $\widetilde{H}^{\textrm{high}}$
is indistinguishable from the unitarity scenario.
Moreover, this result
holds for an arbitrary potential $V\left(x\right)$ since one can
always split $x$ into intervals small enough that $V\left(x\right)$
is constant and then construct the full solution as in eq. (\ref{eq:S_x_dependent_potential}).
So even in matter, for the high scale nonunitarity scenario, nonunitarity effect is proportional to $\left(UU^{\dagger}\right)_{\alpha\beta}$
for $\alpha\neq\beta$ as encapsulated
in eq. (\ref{eq:general_identity}).

In low scale nonunitarity scenario, $\overline{\mathbf{U}}^{-1}=\mathbf{U}^{\dagger}$,
eq. (\ref{eq:H_mass_basis}) becomes
\begin{eqnarray}
\widetilde{H}^{\textrm{low}} & = & \Delta+\mathbf{U}^{\dagger}V\mathbf{U}.
\end{eqnarray}
First of all, since $\mathbf{U}$ is unitary, eqs. (\ref{eq:I_combination}) and (\ref{eq:R_combination})
remain matter invariant as long as $V$ is diagonal and hence the resulting matter invariant identities hold. Next, besides
the two possibilities discussed in the vacuum case, now we have a
new handle: with the matter effect, the difference appear at leading
order in small unitarity violating parameter $\epsilon$ in which the leading Hamiltonian $\widetilde{H}^{\textrm{low}}$ is given by~\cite{Fong:2017gke}
\begin{eqnarray}
\widetilde{H}^{\textrm{low},0} & = & \Delta+\hat{U}^{\dagger}V\hat{U}.
\end{eqnarray}
Comparing the $\widetilde{H}^{\textrm{high}}$ and $\widetilde{H}^{\textrm{low},0}$,
the difference is proportional to
\begin{eqnarray}
\widetilde{H}^{\textrm{low},0}-\widetilde{H}^{\textrm{high}} & = & \hat{U}^{\dagger}V\hat{U}-U^{-1}\kappa V\kappa U^{\dagger,-1},
\label{eq:H_difference_matter}
\end{eqnarray}
where we have written $\overline{U}=\kappa^{-1}U$ with $\kappa\equiv\textrm{diag}\left(\sqrt{\left(UU^{\dagger}\right)_{ee}},\sqrt{\left(UU^{\dagger}\right)_{\mu\mu}},\sqrt{\left(UU^{\dagger}\right)_{\tau\tau}}\right)$. 
Notice that the mapping of $\kappa^{-1} U$ to $\hat U$ in the vacuum case does not work in the presence of matter i.e. eq.~\eqref{eq:H_difference_matter} does not become zero.

Let us suppose that $\hat{U}=U$, i.e., the nonunitarity effect results
in an identical $3\times3$ submatrix. Then $\widetilde{H}^{\textrm{low},0}-\widetilde{H}^{\textrm{high}}\neq0$
and matter effects will result in different eigenvalues and eigenvectors.
For $0\leq x<x_{1}$ where $V$ is constant, we can solve for the
oscillation amplitude $S$ in the flavor basis (\ref{eq:S_constant_V})
as follows
\begin{eqnarray}
S_{\beta\alpha}^{\textrm{high}} & = & \kappa_{\beta}^{-1}\kappa_{\alpha}^{-1}\sum_{i,j,k}U_{\beta i}U_{\alpha j}^{*}X_{ik}^{\textrm{high}}X_{jk}^{\textrm{high},*}e^{-i\lambda_{k}^{\textrm{high}}x},\\
S_{\beta\alpha}^{\textrm{low},0} & = & \sum_{i,j,k}U_{\beta i}U_{\alpha j}^{*}X_{ik}^{\textrm{low}}X_{jk}^{\textrm{low},*}e^{-i\lambda_{k}^{\textrm{low}}x}.
\end{eqnarray}
Hence, besides the amplitudes, the frequencies are different due to different eigenvalues $\lambda_{k}$.

To minimize the difference in matter potential, let us make a different choice $\hat{U}=\kappa U^{\dagger,-1}$ such that $\widetilde{H}^{\textrm{low},0}=\widetilde{H}^{\textrm{high}}$. 
In this case, $\widetilde{H}^{\textrm{high}}$
and $\widetilde{H}^{\textrm{low},0}$ will have exactly the same eigenvalues
$\lambda_{k}$ and are diagonalized by the same unitary matrix $X$. So, we have
\begin{eqnarray}
S_{\beta\alpha}^{\textrm{high}} & = & \kappa_{\beta}^{-1}\kappa_{\alpha}^{-1}\sum_{i,j,k}U_{\beta i}U_{\alpha j}^{*}X_{ik}X_{jk}^{*}e^{-i\lambda_{k}x},\\
S_{\beta\alpha}^{\textrm{low},0} & = & \kappa_{\beta}\kappa_{\alpha}\sum_{i,j,k}\left(U^{\dagger,-1}\right)_{\beta i}\left(U^{\dagger,-1}\right)_{\alpha j}^{*}X_{ik}X_{jk}^{*}e^{-i\lambda_{k}x},
\end{eqnarray}
where the differences are only in the amplitudes. 
Further difference between high scale and low scale nonunitarity scenarios comes from $(\hat P_{\beta})_{\xi\lambda}$ in eq.~\eqref{eq:prob} just like in the vacuum case.

\begin{figure}
\begin{centering}
\includegraphics[scale=0.65]{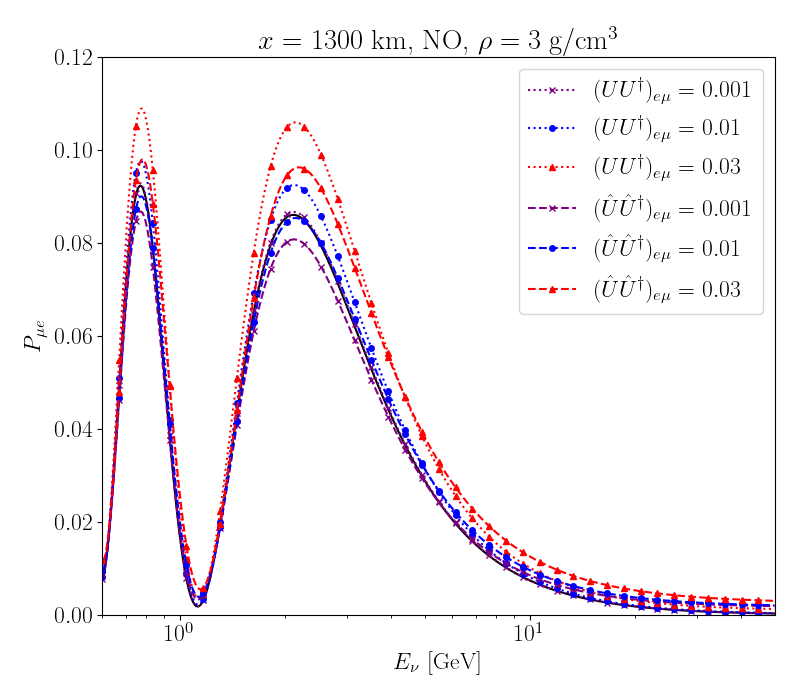}
\par\end{centering}
\caption{Same as Figure \ref{fig:nonunitarity_vacuum} but in a constant matter density of 3 g/cm$^3$.\label{fig:nonunitarity_constant_matter}}
\end{figure}

\begin{figure}
\begin{centering}
\includegraphics[scale=0.4]{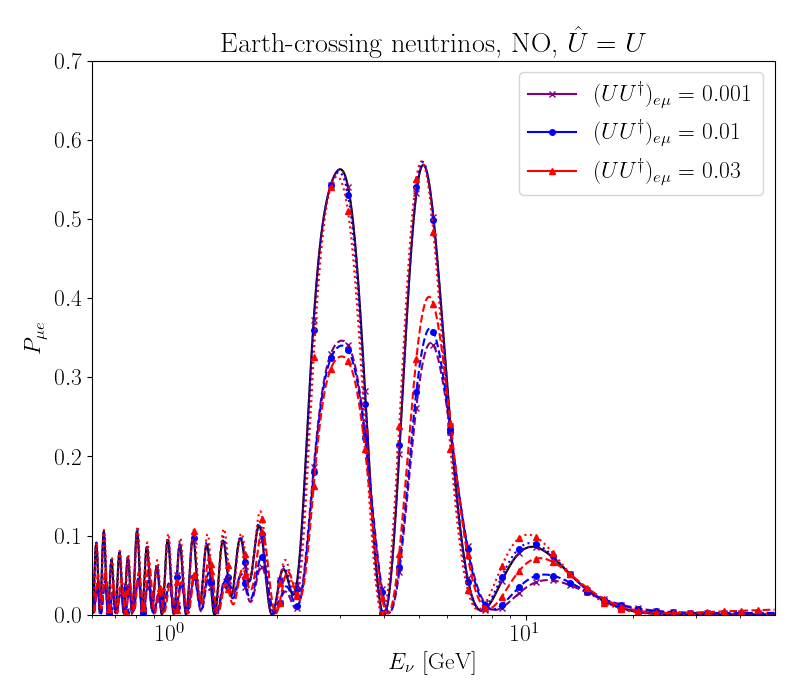}
\includegraphics[scale=0.4]{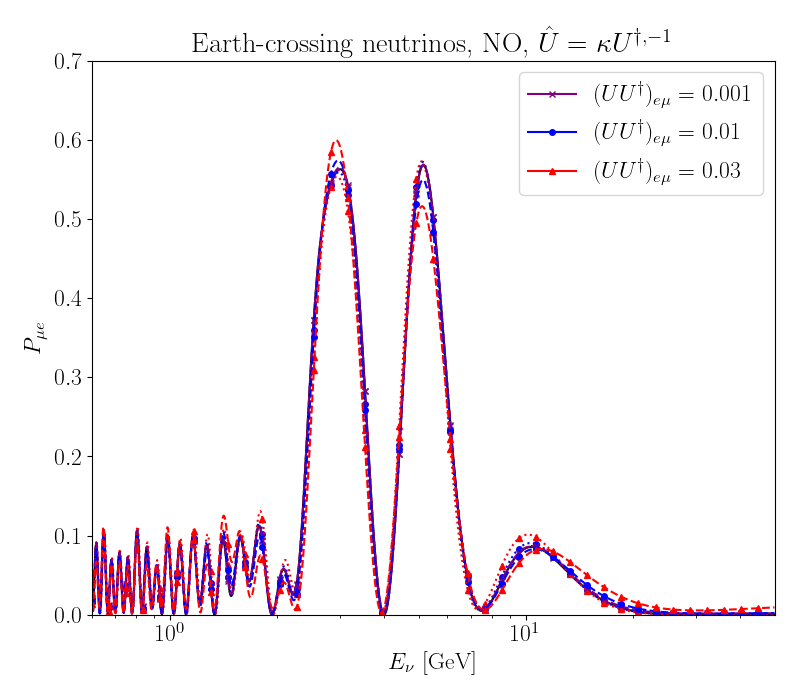}
\par\end{centering}
\caption{Comparison of the probability of $\nu_\mu \to \nu_e$ as a function of neutrino energy $E_\nu$ using the simplified PREM model for neutrino crossing through the Earth core between high scale nonunitarity scenario (dotted lines) and low scale nonunitarity scenario (dashed lines). We have shown the two extreme choices $\hat U = U$ (left plot) and $\hat U = \kappa U^{\dagger,-1}$ (right plot). See text for further explanation. The solid black line is the standard three-flavor unitarity scenario.
	\label{fig:nonunitarity_varying_matter}}
\end{figure}

In Figure \ref{fig:nonunitarity_constant_matter}, we show identical
situation with Figure \ref{fig:nonunitarity_vacuum} except with a constant matter density of $3$ g/cm$^{3}$. Overall, the matter
effect enhances the differences between the high scale and low scale
nonunitarity scenarios as expected. We also observe that with decreasing modulus of
$\left(UU^{\dagger}\right)_{e\mu}$, the high scale nonunitarity scenario
approaches the unitarity scenario (the black line) while this does not
happen for the low scale nonunitarity scenario. In Figure \ref{fig:nonunitarity_varying_matter},
we consider earth core-crossing  neutrinos in a simplified
PREM model (see appendix A of \cite{Fong:2022oim}) with high scale nonunitarity parameters $\left(UU^{\dagger}\right)_{ee}=\left(UU^{\dagger}\right)_{\mu\mu}=0.96$, $\left(UU^{\dagger}\right)_{\tau\tau} = 1$, $\left(UU^{\dagger}\right)_{e\tau} = \left(UU^{\dagger}\right)_{\mu\tau} = 0$ and $\left(UU^{\dagger}\right)_{e\mu}=\left\{ 10^{-3},10^{-2},0.03\right\}$. For low scale nonunitarity parameters, on the left plot, we set $\hat{U}=U$
while on the right plot, we set $\hat{U}=\kappa U^{\dagger,-1}$.
These represent the two extreme cases where for $\hat{U}=U$, the
difference in matter between the two scenarios is maximal while for $\hat{U}=\kappa U^{\dagger,-1}$, the matter effect is identical.

\section{Conclusions}\label{sec:conclusions}

In this work, we have derived analytical oscillation probability
amplitude for arbitrary flavors of neutrinos without assuming a unitary
$\mathbf{U}$. 
With this result, we have proven a theorem that as
long as $\left(\mathbf{U}\mathbf{U}^{\dagger}\right)_{\beta\alpha}=0$
for all $\alpha\neq\beta$, the scenario is indistinguishable from
a unitarity scenario in an arbitrary matter potential. We further derive
a general identity (\ref{eq:general_identity}) which reduces to (\ref{eq:general_identity_vacuum})
in the vacuum and vanishes in the unitarity scenario. 

We have highlighted
the differences between high scale and low scale nonunitarity scenarios
in neutrino oscillations, which are to be expected since in the former case,
all new physics are integrated out while in the latter, new states
are accessible though not necessarily remain coherent to result in
oscillations. 
The first difference is that there is a zero distance effect for high scale nonunitarity scenario due to nonorthogonal flavor states, while it is absence for low scale nonunitarity scenario.
On the one hand, although high scale nonunitarity scenario
is model-independent (all the effects are fully captured by a nonunitary
$U$), nonunitarity effects are proportional to $\left(UU^{\dagger}\right)_{\beta\alpha}$
for $\alpha\neq\beta$ and will be suppressed accordingly if the off-diagonal elements are small. 
On the other hand, while low scale nonunitarity scenario
is model-dependent (depending on the properties of the new states),
an almost model-independent scenario can be obtained if oscillations
involving the new states can be averaged out and nonunitarity effects
can be captured by a nonunitary $\hat{U}$ and a leaking term ${\cal C}_{\alpha\beta}$. In this case, low scale nonunitarity
effects remain even in the limit of vanishing $(\hat{U}\hat{U}^{\dagger})_{\beta\alpha}$
for all $\alpha\neq\beta$. 

The next important result for high scale nonunitarity is that despite $U$ being nonunitary due to new physics, we have shown that the theory remains unitary since the Hamiltonian in the vacuum mass basis is Hermitian and is related to the non-Hermitian Hamiltonian in the flavor basis through similarity transformation with the nonunitary $U$. In order words, unitarity violation is only apparent but we will continue to call this scenario high scale unitarity violation in reference to nonunitary $U$. We have constructed explicitly neutrino oscillation probability which always respect unitarity even though $U$ is not unitary. 

In summary, high scale and low scale nonunitarity
scenarios are distinct, testing them carefully in the neutrino oscillation experiments will certainly give clues to what lies beyond the SM.

\section{Acknowledgments}
C.S.F. acknowledges the support by grant 2019/11197-6 and 2022/00404-3 from São Paulo Research Foundation (FAPESP), and grant 301271/2019-4 and 407149/2021-0 from National Council for Scientific and Technological Development (CNPq). He would like to thank Hisakazu Minakata and Celso Nishi for reading and commenting on the manuscript. He also acknowledges support from the ICTP through the Associates Programme (2023-2028) while the revised version of this work was being completed.

\appendix

\section{Nonorthogonal basis\label{app:nonorthogonal}}

Let us expand in an arbitrary state $\left| \psi \right\rangle$ in a nonorthogonal but normalized basis $\{\left|  \alpha \right\rangle\}$
\begin{eqnarray}
	\left| \psi \right\rangle &=&
	\sum_\alpha c_\alpha \left|  \alpha \right\rangle .
	\label{eq:psi}
\end{eqnarray}
Multiplying the above by $\left\langle  \beta \right|$ and define ${\cal N}_{\beta\alpha} \equiv \left\langle \beta|\alpha\right\rangle$, we can solve for $c_\alpha$ as follows
\begin{eqnarray}
	c_\alpha &=&
	\sum_\beta ({\cal N}^{-1})_{\alpha\beta} \left\langle \beta|\psi\right\rangle .
\end{eqnarray}
Substituting the solution above back to into eq.~\eqref{eq:psi}, we obtain the completeness relation
\begin{eqnarray}
	\sum_{\alpha,\beta}  \left|\alpha\right\rangle ({\cal N}^{-1})_{\alpha\beta} \left\langle \beta \right| &=& \mathbf{I}.
	\label{eq:completeness_nonorthogonal}
\end{eqnarray}
Formally, we can take $g^{\alpha\beta} \equiv ({\cal N}^{-1})_{\alpha\beta}$ as the metric which raises the indices of $\left|_\alpha\right\rangle \equiv \left|\alpha\right\rangle$ as $\left|^\alpha\right\rangle = g^{\alpha\beta} \left|_\beta\right\rangle$ to form the dual vector but we will not use this notation in this work.
Applying the result above to case of neutrino flavor eigenstates which are not orthogonal in general, we identify ${\cal N} = \overline{\mathbf U} \,\overline{\mathbf U}^\dagger$.

\section{Projected probability operator\label{app:probability_operator}}

In refs.~\cite{Leon:1988,Manning:1990}, the theory of projected probabilities on nonorthogonal states are developed. The basic idea is to first
project $\left|\psi\right\rangle $ to a chosen $\left|\alpha\right\rangle $
and to the corresponding orthogonal component $\left|\alpha\right\rangle _{\perp}$.
Then the orthogonal component is further projected to the (hyper)plane
formed by the rest of the basis states and to the orthogonal component
to this (hyper)plane. And this new orthogonal component is again projected to $\left|\alpha\right\rangle $ and $\left|\alpha\right\rangle _{\perp}$
and so on. The procedure can be written down as follows
\begin{eqnarray}
	\left|\psi\right\rangle  & = & \frac{1}{2}\left(p_{\alpha}\left|\psi\right\rangle +q_{\alpha}\left|\psi\right\rangle \right)+\frac{1}{2}\left(p_{\not\alpha}\left|\psi\right\rangle +q_{\not\alpha}\left|\psi\right\rangle \right)\nonumber \\
	& = & \frac{1}{2}\left(p_{\alpha}\left|\psi\right\rangle +p_{\not\alpha}q_{\alpha}\left|\psi\right\rangle +q_{\not\alpha}q_{\alpha}\left|\psi\right\rangle \right)
	 +\frac{1}{2}\left(p_{\not\alpha}\left|\psi\right\rangle +p_{\alpha}q_{\not\alpha}\left|\psi\right\rangle +q_{\alpha}q_{\not\alpha}\left|\psi\right\rangle \right)\nonumber \\
	& = & \frac{1}{2}\left(p_{\alpha}\left|\psi\right\rangle +p_{\not\alpha}q_{\alpha}\left|\psi\right\rangle +p_{\alpha}q_{\not\alpha}q_{\alpha}\left|\psi\right\rangle +...\right)
	+\frac{1}{2}\left(p_{\not\alpha}\left|\psi\right\rangle +p_{\alpha}q_{\not\alpha}\left|\psi\right\rangle +p_{\not\alpha}q_{\alpha}q_{\not\alpha}\left|\psi\right\rangle +...\right)\nonumber \\
	& = & \frac{1}{2}\left[p_{\alpha}\left(\mathbf{I}+q_{\not\alpha}+q_{\not\alpha}q_{\alpha}+...\right)\right]\left|\psi\right\rangle 
	 +\frac{1}{2}\left[p_{\not\alpha}\left(\mathbf{I}+q_{\alpha}+q_{\alpha}q_{\not\alpha}+...\right)\right]\left|\psi\right\rangle ,\label{eq:procedure}
\end{eqnarray}
where 
\begin{eqnarray}
	p_{\alpha} & \equiv & \left|\alpha\right\rangle \left\langle \alpha\right|,\;\;q_{\alpha}\equiv\mathbf{I}-p_{\alpha},
\end{eqnarray}
and $p_{\not\alpha}$ is the projection on to the hyperplane spanned
by the rest of bases besides $\left|\alpha\right\rangle $ and $q_{\not\alpha}\equiv\mathbf{I}-p_{\not\alpha}$.
From the above, we have
\begin{eqnarray}
	P_{\alpha}\left|\psi\right\rangle  & = & c_{\alpha}\left|\alpha\right\rangle =\frac{1}{2}\left[p_{\alpha}\left(\mathbf{I}+q_{\not\alpha}+q_{\not\alpha}q_{\alpha}+...\right)\right]\left|\psi\right\rangle .
\end{eqnarray}
By construction, the sum of absolute square of each operator is unity
\begin{eqnarray}
	\frac{1}{2}\left(p_{\alpha}+q_{\not\alpha}p_{\alpha}q_{\not\alpha}+q_{\alpha}q_{\not\alpha}p_{\alpha}q_{\not\alpha}q_{\alpha}+...\right)+\frac{1}{2}\left(p_{\not\alpha}+q_{\alpha}p_{\not\alpha}q_{\alpha}+q_{\not\alpha}q_{\alpha}p_{\not\alpha}q_{\alpha}q_{\not\alpha}+...\right) & = & \mathbf{I}.\label{eq:sum_operator_squared}
\end{eqnarray}
Hence, we can define the \emph{probability operator} of measuring
$\left|\alpha\right\rangle $ in $\left|\psi\right\rangle $ expanded
in the basis $\left|\alpha\right\rangle $
\begin{eqnarray}
	\hat{P}_{\alpha\alpha} & \equiv & \frac{1}{2}\left(p_{\alpha}+q_{\not\alpha}p_{\alpha}q_{\not\alpha}+q_{\alpha}q_{\not\alpha}p_{\alpha}q_{\not\alpha}q_{\alpha}+...\right).\label{eq:P_aa}
\end{eqnarray}
The second operator can be further decomposed as projection onto $\left|\beta\right\rangle $
and $\left|{\not\beta}\right\rangle $ where $\beta,{\not\beta}\subset{\not\alpha}$
to construct $\hat{P}_{\beta\alpha}$. For $n\leq3$, we can write
down the closed form of probability operators.

\subsection{Two-state system}

For $n=2$, $\alpha,{\not\alpha}=\left\{ 1,2\right\} $, we have only
two operators $\hat{P}_{11}$ and $\hat{P}_{\not22}$ corresponding
to the two terms in eq. (\ref{eq:sum_operator_squared}). Utilizing
the identities $q_{2}q_{1}q_{2}\left|1\right\rangle =\left|{\cal N}_{12}\right|^{2}q_{2}\left|1\right\rangle $
and $q_{1}q_{2}q_{1}\left|2\right\rangle =\left|{\cal N}_{12}\right|^{2}q_{1}\left|2\right\rangle $,
we have a converging geometric series with $\left|N_{12}\right|^{2}<1$
and eq. (\ref{eq:P_aa}) becomes\footnote{In this case, $\alpha\supset\left\{ 1,2\right\} $ and we define this
	new notation such that it is in agreement with the notation we will
	use later for $n=3$.}
\begin{eqnarray}
	\left(p_{\alpha}\right)_{\left\{ 1,2\right\} } & \equiv & \hat{P}_{\alpha\alpha}=\frac{1}{2}\left(p_{\alpha}+\frac{q_{\not\alpha}p_{\alpha}q_{\not\alpha}+q_{\alpha}q_{\not\alpha}p_{\alpha}q_{\not\alpha}q_{\alpha}}{1-\left|{\cal N}_{12}\right|^{4}}\right),
\end{eqnarray}
where the subscript $\left\{ 1,2\right\} $ denotes the basis states.
So $\left(p_{\alpha}\right)_{\left\{ 1,2\right\} }$ is the \emph{probability
	operator} which gives the probability of a state $\left|\psi\right\rangle $
being found along $\left|\alpha\right\rangle $ given by $\left\langle \psi\right|\left(p_{\alpha}\right)_{\left\{ 1,2\right\} }\left|\psi\right\rangle $. After some algebra, we arrive at
\begin{eqnarray}
	\left(p_{\alpha}\right)_{\left\{ 1,2\right\} } & = & \frac{1}{1-\left|{\cal N}_{12}\right|^{4}}\left[p_{\alpha}+\left|{\cal N}_{12}\right|^{2}p_{\not\alpha}-\frac{1+\left|{\cal N}_{12}\right|^{2}}{2}\left\{ p_{\alpha},p_{\not\alpha}\right\} \right],\label{eq:probability_operator_2flavor}
\end{eqnarray}
where $\left\{ p_{\alpha},p_{\not\alpha}\right\} =p_{\alpha}p_{\not\alpha}+p_{\not\alpha}p_{\alpha}$.
Summing over $\alpha$, one obtain
\begin{eqnarray}
	\sum_{\alpha=1}^{2}\left(p_{\alpha}\right)_{\left\{ 1,2\right\} } & = & \frac{p_{1}+p_{2}-\left\{ p_{1},p_{2}\right\} }{1-\left|{\cal N}_{12}\right|^{2}}=\mathbf{I},
\end{eqnarray}
which follows from eq.~\eqref{eq:completeness_nonorthogonal}. 

The probability of $\left|\psi\right\rangle \to\left|\alpha\right\rangle $
is then given by
\begin{eqnarray}
	P_{\psi\to\alpha} & = & \left\langle \psi\right|\left(p_{\alpha}\right)_{\left\{ 1,2\right\} }\left|\psi\right\rangle \nonumber \\
	& = & \sum_{\xi,\zeta,\eta,\lambda}\left\langle \psi|\xi\right\rangle \left({\cal N}^{-1}\right)_{\xi\zeta}\left\langle \zeta\right|\left(p_{\alpha}\right)_{\left\{ 1,2\right\} }\left|\eta\right\rangle \left({\cal N}^{-1}\right)_{\eta\lambda}\left\langle \lambda|\psi\right\rangle \nonumber \\
	& \equiv & \left(S^{\dagger}D_{\alpha}S\right)_{\psi\psi},
\end{eqnarray}
where we have defined
\begin{eqnarray}
	S_{\lambda\psi} & \equiv & \left\langle \lambda|\psi\right\rangle ,\\
	\left[D_{\alpha}\right]_{\xi\lambda} & \equiv & \sum_{\zeta,\eta}\left({\cal N}^{-1}\right)_{\xi\zeta}\left\langle \zeta\right|\left(p_{\alpha}\right)_{12}\left|\eta\right\rangle \left({\cal N}^{-1}\right)_{\eta\lambda}.
\end{eqnarray}
In the second equality of the probability, we have used the completeness relation given by eq.~\eqref{eq:completeness_nonorthogonal}. For the two-flavor system, we have~
\begin{eqnarray}
	{\cal N} & = & \left(\begin{array}{cc}
		1 & {\cal N}_{12}\\
		{\cal N}_{12}^{*} & 1
	\end{array}\right),\quad {\cal N}^{-1}=\frac{1}{1-\left|{\cal N}_{12}\right|^{2}}\left(\begin{array}{cc}
		1 & -{\cal N}_{12}\\
		-{\cal N}_{12}^{*} & 1
	\end{array}\right).
\end{eqnarray}
Evaluating directly the matrix elements of $D_{\alpha}$, we have
\begin{eqnarray}
	\left[D_{\alpha}\right]_{\xi\lambda} & = & \frac{1}{1-\left|{\cal N}_{12}\right|^{4}}\bigg[\delta_{\xi\alpha}\delta_{\alpha\lambda}+\left|{\cal N}_{12}\right|^{2}\delta_{\xi\not\alpha}\delta_{\not\alpha\lambda}\nonumber \\
	&& -\frac{1+\left|{\cal N}_{12}\right|^{2}}{2}\left(\delta_{\xi\alpha}{\cal N}_{\alpha\not\alpha}\delta_{\not\alpha\lambda}+\delta_{\xi\not\alpha}{\cal N}_{\not\alpha\alpha}\delta_{\alpha\lambda}\right)\bigg],
\end{eqnarray}
or
\begin{eqnarray}
	\left[D_{\alpha}\right]_{\xi\lambda} & = & \begin{cases}
		{\displaystyle 1 + \frac{\left|{\cal N}_{12}\right|^{4}}{1-\left|{\cal N}_{12}\right|^{4}}}, & \xi=\lambda=\alpha\\
		{\displaystyle \frac{\left|{\cal N}_{12}\right|^{2}}{1-\left|{\cal N}_{12}\right|^{4}}}, & \xi=\lambda\neq\alpha\\
		{\displaystyle-\frac{1}{2}\frac{{\cal N}_{\xi\lambda}}{1-\left|{\cal N}_{12}\right|^{2}}}, & \xi\neq\lambda\;\textrm{and}\;\xi=\alpha\;\textrm{or}\;\lambda=\alpha
	\end{cases}.
\end{eqnarray}
In the absence of nonorthogonality, the standard result is recovered
$\left(p_{\alpha}\right)_{\left\{ 1,2\right\} }=p_{\alpha}$, $\left[D_{\alpha}\right]_{\xi\lambda}=\delta_{\xi\alpha}\delta_{\alpha\lambda}$
and $P_{\psi\to\alpha}=\left|S_{\alpha\psi}\right|^{2}$.

\subsection{Three-state system}

For $n=3$, in order to obtain symmetrize probability operator, one repeats the procedure of eq. (\ref{eq:procedure}) projecting $\left|\psi\right\rangle $
into $\left|\beta\right\rangle \neq\left|\alpha\right\rangle $ and
the corresponding plane not containing $\left|\beta\right\rangle $
and there are altogether 3 choices. Doing so, we obtain the symmetrize
probability operator on $\left|\alpha\right\rangle $ as
\begin{eqnarray}
	\hat{P}_{\alpha} & = & \frac{1}{3}\sum_{\beta}\hat{P}_{\alpha\beta},
\end{eqnarray}
where
\begin{eqnarray}
	\hat{P}_{\alpha\alpha} & = & \frac{1}{2}\left[p_{\alpha}+\frac{q_{\beta\gamma}p_{\alpha}q_{\beta\gamma}}{1-X_{\alpha}^{2}}+\frac{q_{\alpha}q_{\beta\gamma}p_{\alpha}q_{\beta\gamma}q_{\alpha}}{1-X_{\alpha}^{2}}\right],\quad\beta\neq\gamma;\beta,\gamma\neq\alpha,\\
	\hat{P}_{\alpha\beta} & = & \frac{1}{2}\Biggl[\left(p_{\alpha}\right)_{\left\{ \alpha,\gamma\right\} }+q_{\beta}\left(p_{\alpha}\right)_{\left\{ \alpha,\gamma\right\} }q_{\beta} \\
	&  & \!\!\!\!\!+\frac{q_{\alpha\gamma}p_{\beta}\left(p_{\alpha}\right)_{\left\{ \alpha,\gamma\right\} }p_{\beta}q_{\alpha\gamma}}{1-X_{\beta}^{2}}+\frac{q_{\beta}q_{\alpha\gamma}p_{\beta}\left(p_{\alpha}\right)_{\left\{ \alpha,\gamma\right\} }p_{\beta}q_{\alpha\gamma}q_{\beta}}{1-X_{\beta}^{2}}\Biggr],\quad\beta\neq\alpha;\gamma\neq\left\{ \alpha,\beta\right\} ,
\end{eqnarray}
with
\begin{eqnarray}
	p_{\beta\gamma} & \equiv & \frac{p_{\beta}+p_{\gamma}-\left\{ p_{\beta},p_{\gamma}\right\} }{1-\left|{\cal N}_{\beta\gamma}\right|^{2}},\;\;q_{\beta\gamma}\equiv\mathbf{I}-p_{\beta\gamma},\;\;\;\gamma\neq\beta,\\
	\left(p_{\alpha}\right)_{\left\{ \alpha,\gamma\right\} } & \equiv & \frac{1}{1-\left|{\cal N}_{\alpha\gamma}\right|^{4}}\left[p_{\alpha}+\left|{\cal N}_{\alpha\gamma}\right|^{2}p_{\not\alpha}-\frac{1+\left|{\cal N}_{\alpha\gamma}\right|^{2}}{2}\left\{ p_{\alpha},p_{\not\alpha}\right\} \right],\quad\gamma\neq\alpha,\\
	X_{\alpha} & \equiv & 1-\frac{\det {\cal N}}{\det {\cal N}_{\alpha}},
\end{eqnarray}
In the last relation above, ${\cal N}_{\alpha}$ is the $2\times2$ matrix
constructed excluding the basis $\left|\alpha\right\rangle $. The
operator $p_{\beta\gamma}$ projects a state onto the $\left\{ \beta,\gamma\right\} $-plane
spanned by two orthonormal vectors $\left|\beta\right\rangle $ and
$\left(1-{\cal N}_{\beta\gamma}^{2}\right)^{-1/2}q_{\beta}\left|\gamma\right\rangle $.
The second operator $\left(p_{\alpha}\right)_{\left\{ \alpha,\gamma\right\} }$
has the same form as the probability operator (\ref{eq:probability_operator_2flavor})
that we have found in the two-flavor case. For the three-flavor case, we can write
\begin{eqnarray}
	{\cal N} & = & \left(\begin{array}{ccc}
		1 & {\cal N}_{12} & {\cal N}_{13}\\
		{\cal N}_{12}^{*} & 1 & {\cal N}_{23}\\
		{\cal N}_{13}^{*} & {\cal N}_{23}^{*} & 1
	\end{array}\right), \\
	{\cal N}^{-1} & = & \frac{1}{\det {\cal N}}\left(\begin{array}{ccc}
		1-\left|{\cal N}_{23}\right|^{2} & -{\cal N}_{12}+{\cal N}_{13}{\cal N}_{23}^{*} & -{\cal N}_{13}+{\cal N}_{12}{\cal N}_{23}\\
		-{\cal N}_{12}^{*}+{\cal N}_{13}^{*}{\cal N}_{23} & 1-\left|{\cal N}_{13}\right|^{2} & -{\cal N}_{23}+{\cal N}_{13}{\cal N}_{12}^{*}\\
		-{\cal N}_{13}^{*}+{\cal N}_{12}^{*}{\cal {\cal N}}_{23}^{*} & -{\cal N}_{23}^{*}+{\cal N}_{13}^{*}{\cal N}_{12} & 1-\left|{\cal N}_{12}\right|^{2}
	\end{array}\right), \\
	\det {\cal N} & = & -\left|{\cal N}_{23}\right|^{2}-\left|{\cal N}_{12}\right|^{2}-\left|{\cal N}_{13}\right|^{2}+{\cal N}_{12}{\cal N}_{13}^{*}{\cal N}_{23}+{\cal N}_{13}{\cal N}_{12}^{*}{\cal N}_{23}^{*},\\
	\det {\cal N}_{1} & = & 1-\left|{\cal N}_{23}\right|^{2},\;\det {\cal N}_{2}=1-\left|{\cal N}_{13}\right|^{2},\;\det {\cal N}_{3}=1-\left|{\cal N}_{12}\right|^{2}.
\end{eqnarray}
One can check explicitly that
\begin{eqnarray}
	\sum_{\alpha=1}^{3}\hat{P}_{\alpha} & = & \frac{1}{3}\sum_{\alpha,\beta}\hat{P}_{\alpha\beta}=\frac{1}{\det {\cal N}}\sum_{\alpha}\left[p_{\alpha}\det {\cal N}_{\alpha}+\sum_{\beta\neq\alpha}\left(-p_{\alpha}p_{\beta}+p_{\alpha}p_{\gamma}p_{\beta}\right)\right]=\mathbf{I},
\end{eqnarray}
which follows from eq.~\eqref{eq:completeness_nonorthogonal}. 

The probability of $\left|\psi\right\rangle \to\left|\alpha\right\rangle $
now becomes
\begin{eqnarray}
	P_{\psi\to\alpha} & = & \langle \psi|\hat{P}_{\alpha}|\psi\rangle \equiv\frac{1}{3}\left[S^{\dagger}\left(E_{\alpha}+\sum_{\beta\neq\alpha}F_{\alpha\beta}\right)S\right]_{\psi\psi},
\end{eqnarray}
where we have defined
\begin{eqnarray}
	\left[E_{\alpha}\right]_{\xi\lambda} & \equiv & \sum_{\zeta,\eta}\left({\cal N}^{-1}\right)_{\xi\zeta}\left\langle \zeta\right|\hat{P}_{\alpha\alpha}\left|\eta\right\rangle \left({\cal N}^{-1}\right)_{\eta\lambda},\\
	\left[F_{\alpha\beta}\right]_{\xi\lambda} & \equiv & \sum_{\zeta,\eta}\left({\cal N}^{-1}\right)_{\xi\zeta}\left\langle \zeta\right|\hat{P}_{\alpha\beta}\left|\eta\right\rangle \left({\cal N}^{-1}\right)_{\eta\lambda}.
\end{eqnarray}
Evaluating explicitly the matrix elements, we obtain
	\begin{eqnarray}
		\left[E_{\alpha}\right]_{\xi\lambda} & = & \begin{cases}
			{\displaystyle 1 +  \frac{X_{\alpha}^{2}}{1-X_{\alpha}^{2}}}, & \xi=\lambda=\alpha\\
			{\displaystyle \frac{\left|{\cal N}_{\alpha\xi}-{\cal N}_{\alpha\gamma}{\cal N}_{\gamma\xi}\right|^{2}}{\left(\det {\cal N}_{\alpha}\right)^{2}\left(1-X_{\alpha}^{2}\right)},\;\gamma\neq\left\{ \alpha,\xi\right\} }, & \xi=\lambda\neq\alpha\\
			{\displaystyle -\frac{1}{2}\frac{{\cal N}_{\xi\lambda}-{\cal N}_{\xi\gamma}{\cal N}_{\gamma\lambda}}{\det {\cal N}}},\;\gamma\neq\left\{ \alpha,\xi\right\} , & \xi\neq\lambda\;\textrm{and}\;\xi=\alpha\;\textrm{or}\;\lambda=\alpha\\
			{\displaystyle \frac{\left({\cal N}_{\alpha\lambda}-{\cal N}_{\alpha\xi}{\cal N}_{\xi\lambda}\right)\left({\cal N}_{\xi\alpha}-{\cal N}_{\xi\lambda}{\cal N}_{\lambda\alpha}\right)}{\left(\det {\cal N}_{\alpha}\right)^{2}\left(1-X_{\alpha}^{2}\right)}}, & \xi\neq\lambda\;\textrm{and}\;\left\{ \xi,\lambda\right\} \neq\alpha
		\end{cases},
	\end{eqnarray}
	and for $\alpha\neq\beta$ and $\gamma\neq\left\{ \alpha,\beta\right\} $
	\begin{eqnarray}
		\left[F_{\alpha\beta}\right]_{\xi\lambda} & = & \begin{cases}
			{\displaystyle \frac{1}{1-\left|{\cal N}_{\alpha\gamma}\right|^{4}}+\frac{\left\langle \beta\right|\left(p_{\alpha}\right)_{\alpha\gamma}\left|\beta\right\rangle \left|{\cal N}_{\alpha\beta}-{\cal N}_{\alpha\gamma}{\cal N}_{\gamma\beta}\right|^{2}}{\left(\det {\cal N}_{\beta}\right)^{2}\left(1-X_{\beta}^{2}\right)}}, & \xi=\lambda=\alpha\\
			{\displaystyle \begin{split}-\frac{1}{2}\frac{1}{1-\left|{\cal N}_{\alpha\gamma}\right|^{4}}\left({\cal N}_{\xi\lambda}-\frac{1+\left|{\cal N}_{\alpha\gamma}\right|^{2}}{2}{\cal N}_{\xi\gamma}{\cal N}_{\gamma\lambda}\right)\\
					-\frac{1}{2}\frac{\left\langle \beta\right|\left(p_{\alpha}\right)_{\left\{ \alpha,\gamma\right\} }\left|\beta\right\rangle }{1-X_{\beta}^{2}}\frac{{\cal N}_{\xi\lambda}-{\cal N}_{\xi\gamma}{\cal N}_{\gamma\lambda}}{\det {\cal N}_{\beta}}\left(1+\left\langle \beta\right|p_{\alpha\gamma}\left|\beta\right\rangle \right)
				\end{split}
			}, & \xi\lambda=\alpha\beta\;\textrm{or}\;\xi\lambda=\beta\alpha\\
			{\displaystyle -\frac{1}{2}\frac{{\cal N}_{\xi\lambda}}{\det {\cal N}_{\beta}}
			+\frac{\left\langle \beta\right|\left(p_{\alpha}\right)_{\left\{ \alpha,\gamma\right\} }\left|\beta\right\rangle \left({\cal N}_{\xi\beta}-{\cal N}_{\xi\lambda}{\cal N}_{\lambda\beta}\right)\left({\cal N}_{\beta\lambda}-{\cal N}_{\beta\xi}{\cal N}_{\xi\lambda}\right)}{\left(\det {\cal N}_{\beta}\right)^{2}\left(1-X_{\beta}^{2}\right)}}, & \xi\lambda=\alpha\gamma\;\textrm{or}\;\xi\lambda=\gamma\alpha\\
			{\displaystyle \frac{1}{2}\left[1+\frac{1+\left\langle \beta\right|p_{\alpha\gamma}\left|\beta\right\rangle ^{2}}{1-X_{\beta}^{2}}\right]\left\langle \beta\right|\left(p_{\alpha}\right)_{\left\{ \alpha,\gamma\right\} }\left|\beta\right\rangle }, & \xi=\lambda=\beta\\
			{\displaystyle \begin{split}-\frac{1}{2}\frac{1}{1-\left|{\cal N}_{\alpha\gamma}\right|^{4}}\left(\left|{\cal N}_{\alpha\gamma}\right|^{2}{\cal N}_{\xi\lambda}-\frac{1+\left|{\cal N}_{\alpha\gamma}\right|^{2}}{2}{\cal N}_{\xi\alpha}{\cal N}_{\alpha\lambda}\right)\\
					-\frac{1}{2}\frac{\left\langle \beta\right|\left(p_{\alpha}\right)_{\left\{ \alpha,\gamma\right\} }\left|\beta\right\rangle }{1-X_{\beta}^{2}}\frac{{\cal N}_{\xi\lambda}-{\cal N}_{\xi\alpha}{\cal N}_{\alpha\lambda}}{\det {\cal N}_{\beta}}\left(1+\left\langle \beta\right|p_{\alpha\gamma}\left|\beta\right\rangle \right)
				\end{split}
				,} & \xi\lambda=\beta\gamma\;\textrm{or}\;\xi\lambda=\gamma\beta\\
			{\displaystyle \frac{\left|{\cal N}_{\alpha\gamma}\right|^{2}}{1-\left|{\cal N}_{\alpha\gamma}\right|^{4}}+\frac{\left\langle \beta\right|\left(p_{\alpha}\right)_{\left\{ \alpha,\gamma\right\} }\left|\beta\right\rangle \left|{\cal N}_{\gamma\beta}-{\cal N}_{\gamma\alpha}{\cal N}_{\alpha\beta}\right|^{2}}{\left(\det {\cal N}_{\beta}\right)^{2}\left(1-X_{\beta}^{2}\right)}}, & \xi=\lambda=\gamma
		\end{cases},
	\end{eqnarray}
	where
	\begin{eqnarray}
		\left\langle \beta\right|p_{\alpha\gamma}\left|\beta\right\rangle  & = & \frac{\left|{\cal N}_{\alpha\beta}\right|^{2}+\left|{\cal N}_{\beta\gamma}\right|^{2}-2\textrm{Re}\left({\cal N}_{\beta\alpha}{\cal N}_{\alpha\gamma}{\cal N}_{\gamma\beta}\right)}{1-\left|{\cal N}_{\alpha\gamma}\right|^{2}},\\
		\left\langle \beta\right|\left(p_{\alpha}\right)_{\left\{ \alpha,\gamma\right\} }\left|\beta\right\rangle  & = & \frac{\left|{\cal N}_{\alpha\beta}\right|^{2}+\left|{\cal N}_{\alpha\gamma}\right|^{2}\left|{\cal N}_{\beta\gamma}\right|^{2} 
			-\left(1+\left|{\cal N}_{\alpha\gamma}\right|^{2}\right)\textrm{Re}\left({\cal N}_{\beta\alpha}{\cal N}_{\alpha\gamma}{\cal N}_{\gamma\beta}\right)}{1-\left|{\cal N}_{\alpha\gamma}\right|^{4}}.
	\end{eqnarray}

\bibliography{nuprobe}

\end{document}